\documentclass[apj]{emulateapj}

\usepackage{apjfonts}

\newcommand{\lya}{\textrm{Ly}\textsc{$\alpha$}}
\newcommand{\lyalpha}{\textrm{Ly}\textsc{$\alpha$}~}
\newcommand{\nv}{\textrm{N}~\textsc{v}}
\newcommand{\nfive}{\textrm{N}~\textsc{v}~}
\newcommand{\siiv}{\textrm{Si}~\textsc{iv}}
\newcommand{\sifour}{\textrm{Si}~\textsc{iv}~}

\newcommand{\siii}{\textrm{Si}~\textsc{ii}}
\newcommand{\civ}{\textrm{C}~\textsc{iv}}
\newcommand{\cfour}{\textrm{C}~\textsc{iv}~}
\newcommand{\cii}{\textrm{C}~\textsc{ii}}
\newcommand{\mgii}{\textrm{Mg}~\textsc{ii}}
\newcommand{\mgtwo}{\textrm{Mg}~\textsc{ii}~}
\newcommand{\feii}{\textrm{Fe}~\textsc{ii}}
\newcommand{\fetwo}{\textrm{Fe}~\textsc{ii}~}
\newcommand{\othree}{[\textrm{O}~\textsc{iii}]~}

\newcommand{\heii}{\textrm{He}~\textsc{ii}}
\newcommand{\hone}{\textrm{H}~\textsc{i}~}
\newcommand{\lrest}{\lambda_{\scriptsize{\textnormal{rest}}}}

\newcommand{\ewrest}{EW$_{\scriptsize{\textnormal{rest}}}$}
\newcommand{\ew}{\textnormal{EW}}

\shorttitle{High-Redshift Weak Emission-Line Quasars}
\shortauthors{Diamond-Stanic et al.}

\begin{document}
\slugcomment{Published in The Astrophysical Journal}

\title{High-Redshift SDSS Quasars with Weak Emission Lines}

\author{Aleksandar M. Diamond-Stanic\altaffilmark{1}, Xiaohui
Fan\altaffilmark{1,2}, W. N. Brandt\altaffilmark{3}, Ohad
Shemmer\altaffilmark{3,4}, Michael A. Strauss\altaffilmark{5}, Scott
F. Anderson\altaffilmark{6}, Christopher L. Carilli\altaffilmark{7},
Robert R. Gibson\altaffilmark{3}, Linhua Jiang\altaffilmark{1},
J. Serena Kim\altaffilmark{1}, Gordon T. Richards\altaffilmark{8},
Gary D. Schmidt\altaffilmark{1}, Donald P. Schneider\altaffilmark{3},
Yue Shen\altaffilmark{5}, Paul S. Smith\altaffilmark{1}, Marianne
Vestergaard\altaffilmark{1,9}, Jason E. Young\altaffilmark{1,3}}

\altaffiltext{1}{Steward Observatory, University of Arizona,
933 North Cherry Avenue, Tucson, AZ 85721, USA; \\ adiamond@as.arizona.edu} 
\altaffiltext{2}{Max Planck Institute for Astronomy,
D-69117, Heidelberg, Germany}
\altaffiltext{3}{Department of Astronomy and Astrophysics, 
Pennsylvania State University, University Park, PA 16802, USA}
\altaffiltext{4}{Current address: Department of Physics, University of
North Texas, Denton, TX 76203}
\altaffiltext{5}{Princeton University Observatory, Peyton Hall, 
Princeton, NJ 08544, USA}
\altaffiltext{6}{Department of Astronomy, University of Washington, 
Box 351580, Seattle, WA 98195, USA}
\altaffiltext{7}{National Radio Astronomy Observatory, P.O. Box O, 
Socorro, NM 87801, USA}
\altaffiltext{8}{Department of Physics, Drexel University, 3141
Chestnut Street, Philadelphia, PA 19104, USA}
\altaffiltext{9}{Department of Physics and Astronomy, Tufts University,
Robinson Hall, 212 College Avenue, Medford, MA 02155, USA} 

\begin{abstract}
We identify a sample of 74 high-redshift quasars ($z>3$) with weak
emission lines from the Fifth Data Release of the Sloan Digital Sky
Survey and present infrared, optical, and radio observations of a
subsample of four objects at $z>4$.  These weak emission-line quasars
(WLQs) constitute a prominent tail of the \lya+\nfive equivalent width
distribution, and we compare them to quasars with more typical
emission-line properties and to low-redshift active galactic nuclei
with weak/absent emission lines, namely BL Lac objects.  We find that
WLQs exhibit hot ($T\sim1000$~K) thermal dust emission and have
rest-frame 0.1--5~$\mu$m spectral energy distributions that are quite
similar to those of normal quasars.  The variability, polarization,
and radio properties of WLQs are also different from those of BL Lacs,
making continuum boosting by a relativistic jet an unlikely physical
interpretation.  The most probable scenario for WLQs involves
broad-line region properties that are physically distinct from those
of normal quasars.
\end{abstract}


\section{Introduction}\label{sec:intro}
The optical and ultraviolet (UV) spectra of type 1 quasars are
characterized by a blue power-law continuum and strong, broad
permitted emission lines with
$\textnormal{FWHM}=1000$--$20,000$~km~s$^{-1}$.  These lines are
thought to originate in photoionized gas located
$\sim10^{16\textnormal{\scriptsize{--}}18}$ cm from the central black
hole, with differential gas velocities producing Doppler-broadened
line profiles \citep[e.g.,][]{pet97, kro99}.  This gas may be confined
to individual clouds outside of the accretion disk or it may be
related to the accretion disk itself, perhaps as part of a disk wind
\citep[e.g.,][]{emm92, mur95}.

Observationally, the strongest of these emission lines, in terms of
both flux and equivalent width (EW), is \lya~$\lambda1216$.  Results
from UV spectra at $z<2$ and optical spectra at $z>2$
\citep[e.g.,][]{sch91, fra93, osm94, war94, zhe97, bro01, die02} show
that the rest-frame EW of \lya+\nv~$\lambda1240$ is typically
50--110~\AA.  These results, however, are based on small numbers of
sources, with tens to hundreds of quasars per study.  The Sloan
Digital Sky Survey \citep[SDSS,][]{yor00} has substantially increased
the available spectroscopic sample of high-redshift quasars; more than
$5000$ quasars have been discovered at $z>3$ \citep{sch07}, extending
all the way to $z=6.42$ \citep{fan03}.  The SDSS quasar selection
algorithm at these redshifts \citep{ric02a} is based largely on the
red colors produced by the Lyman break ($\lrest=912$~\AA) and the
onset of the \lyalpha forest ($\lrest=1216$~\AA).  It is sensitive to
quasars with bright UV continua, without a strong dependence on
emission-line strength \citep{fan01}, and a small fraction of
high-redshift quasars has been found with very weak or absent \lyalpha
emission \citep[e.g.,][]{fan99, and01, col05, fan06}.

The first weak-emission line quasar (WLQ) found at high redshift was
SDSS J153259.96-003944.1 \citep[$z=4.62$,][hereafter
SDSSJ1532]{fan99}.  Its flat continuum and lack of emission lines
suggested that it could be the highest-redshift BL Lac object ever
found, but its paucity of radio flux, X-ray flux, and optical
polarization were inconsistent with the BL Lac hypothesis.  The
spectra of classical BL Lacs lack emission lines because their
continuum emission is relativistically boosted by a jet along the line
of sight, and they are usually radio-loud, X-ray-loud, and highly
polarized \citep[e.g.,][]{urr95}.  The nature of SDSSJ1532 was left as
an open question.

Subsequently, two additional WLQs, SDSS~J130216.13+003032.1 ($z=4.47$,
hereafter SDSSJ1302) and SDSS~J144231.72+011055.2 ($z=4.51$, hereafter
SDSSJ1442), were identified by \citet{and01}, and the strong radio and
X-ray emission of the latter \citep{sch05, shem06} indicated that its
continuum could be moderately beamed.  In a sample of optically
selected BL Lac candidates, \citet{col05} found seven sources at
$z>2.7$, including three radio sources and one X-ray source
\citep{sch03}.  The highest-redshift WLQ, SDSS~J133550.81+353315.8 at
$z=5.93$, was discovered by \citet{fan06}, who discussed whether it
could be a strongly lensed galaxy, a BL Lac object, or a quasar with a
very weak emission-line region.

Further insight into the nature of WLQs has been gained from pointed
X-ray observations with {\it Chandra}.  \citet{vig01} presented
observations of SDSSJ1532, while \citet{shem06} presented deeper data
for SDSSJ1532, as well as observations of SDSSJ1302, SDSSJ1442, and
SDSS~J140850.91+020522.7 ($z=4.01$, hereafter SDSSJ1408);
\citet{shem06} discussed whether the weak emission lines could be due
to continuum boosting or a deficit of line-emitting gas.
\citet{shem09} presented deeper data for SDSSJ1302 as well as
observations of eight additional WLQs at $z>2.7$; they also compared
the optical, X-ray, and radio properties of WLQs to those of BL Lacs,
and discussed whether WLQs could be extreme quasars with high
accretion rates.  They found that WLQs are weaker than BL Lacs at
radio and X-ray wavelengths relative to the optical, but that there is
no evidence of steep hard X-ray spectra characteristic of high
accretion rates.

This paper extends the study of WLQs to the full high-redshift ($z>3$)
sample that has been discovered by the SDSS, and uses a rich,
multiwavelength data set for a subsample of four WLQs at $z>4$ to test
the available hypotheses regarding their nature.  In
Section~\ref{sec:sample}, we measure EW(\lya+\nv) for each $z>3$
quasar in the Fifth Data Release Quasar Catalog \citep{sch07}, select
quasars with EW values $>3\sigma$ below the mean of the distribution,
and compare their properties to those of the general high-redshift
quasar population.  Sections~\ref{sec:ir}, \ref{sec:optical}, and
\ref{sec:synchrotron} then present results from an extensive
observational campaign focusing on the WLQs SDSSJ1302, SDSSJ1408,
SDSSJ1442, and SDSSJ1532.  These data include {\it Spitzer}
3.6--24~$\mu$m mid-infrared (mid-IR) photometry and near-IR photometry
in $J$, $H$, and $K_{s}$ bands that are compared to the spectral
energy distributions (SEDs) of typical quasars (Section~\ref{sec:ir});
multiple epochs of optical spectroscopy and $i$-band photometry (in
addition to the available SDSS data) that constrain the line and
continuum variability of WLQs (Section~\ref{sec:optical}); and optical
imaging polarimetry along with Very Large Array (VLA) radio continuum
observations at $L$ and $C$ bands that test for the presence of
synchrotron emission from a relativistic jet
(Section~\ref{sec:synchrotron}).  We discuss the implications of our
results and the connection to lower-redshift quasars with weak
emission lines \citep[e.g.,][]{mcd95, lei07a} in
Section~\ref{sec:discussion}, and we summarize and conclude in
Section~\ref{sec:conclusions}.

We assume a cosmology with $\Omega_\Lambda=0.7$, $\Omega_M=0.3$, and
$H_0=70$~km~s$^{-1}$~Mpc$^{-1}$.  All quoted emission-line EWs are in
the rest frame of the quasar.

\begin{figure}
\begin{center}
\includegraphics[angle=0,scale=0.4]{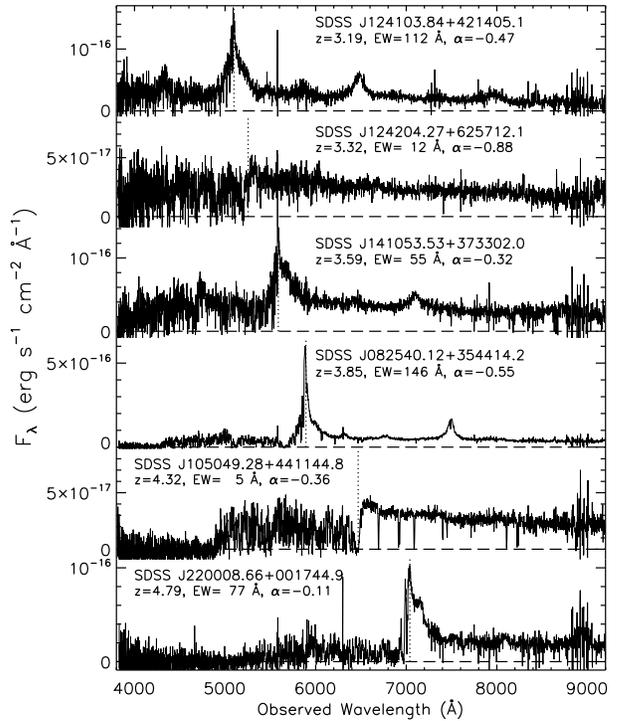}
\caption{Spectra of six $z>3$ quasars exhibiting a range of redshifts,
\lya+\nfive emission-line EWs, continuum spectral slopes, and S/N.
The wavelength corresponding to $\lrest=1216$~\AA\ is marked by a
dotted line in each spectrum.  The objects in the second and fifth
panels are WLQs.}
\label{fig:example_spectra}
\end{center}
\end{figure}

\begin{figure*}
\begin{center}
\includegraphics[angle=90,scale=0.65]{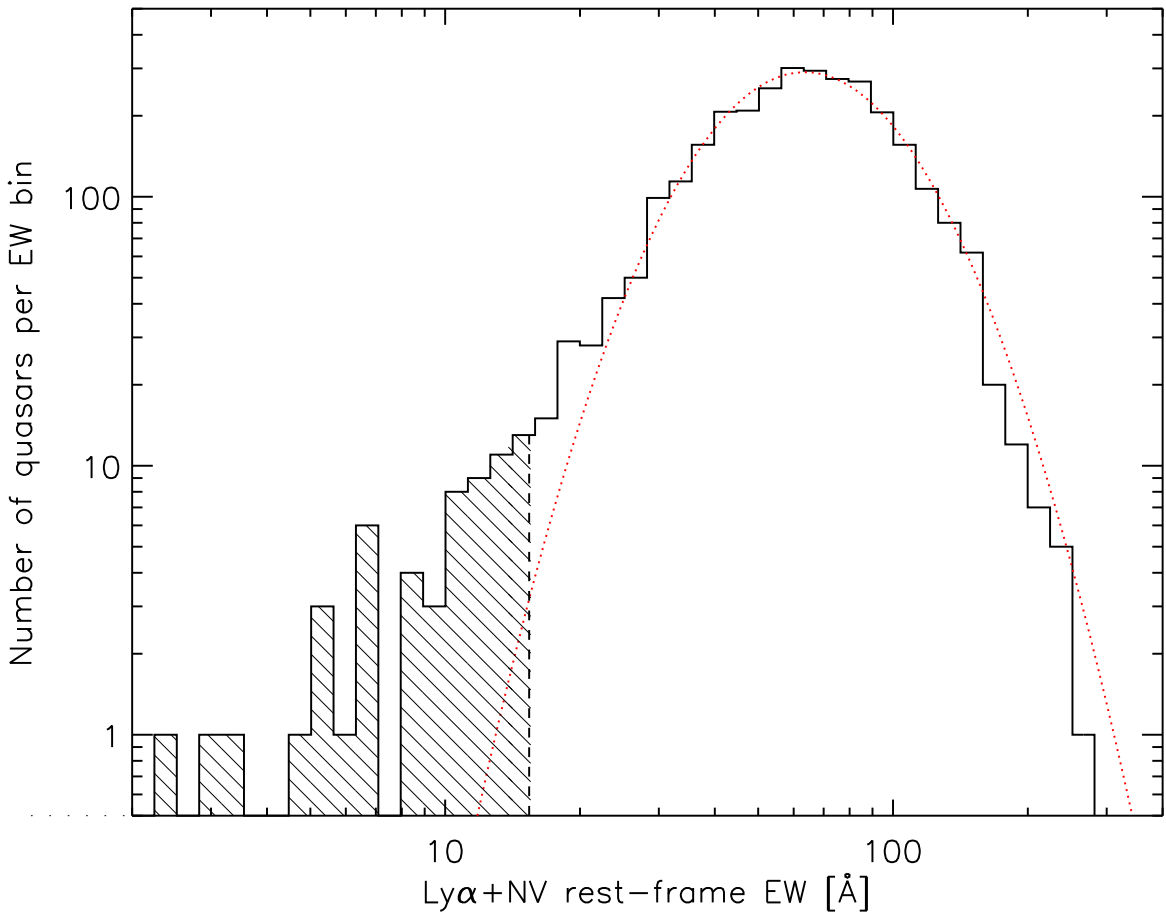}
\caption{Distribution of rest-frame \lya+\nfive EWs for the sample of
3058 $z>3$ non-BAL quasars described in Section~\ref{sec:sample}.  The
dotted red line shows the best-fit log-normal distribution, with a
mean EW of $63.6$~\AA\ and a $1\sigma$ range of $39.6$--$101.9$~\AA.
The fit is a reasonable description of the data above
$\textnormal{EW}=20$~\AA, but there is a prominent tail toward low EW
values.  The dashed line and the hashed section of the histogram
indicate our WLQ definition.}
\label{fig:ew_distribution}
\end{center}
\end{figure*}

\begin{figure}
\begin{center}
\includegraphics[angle=0,scale=0.4]{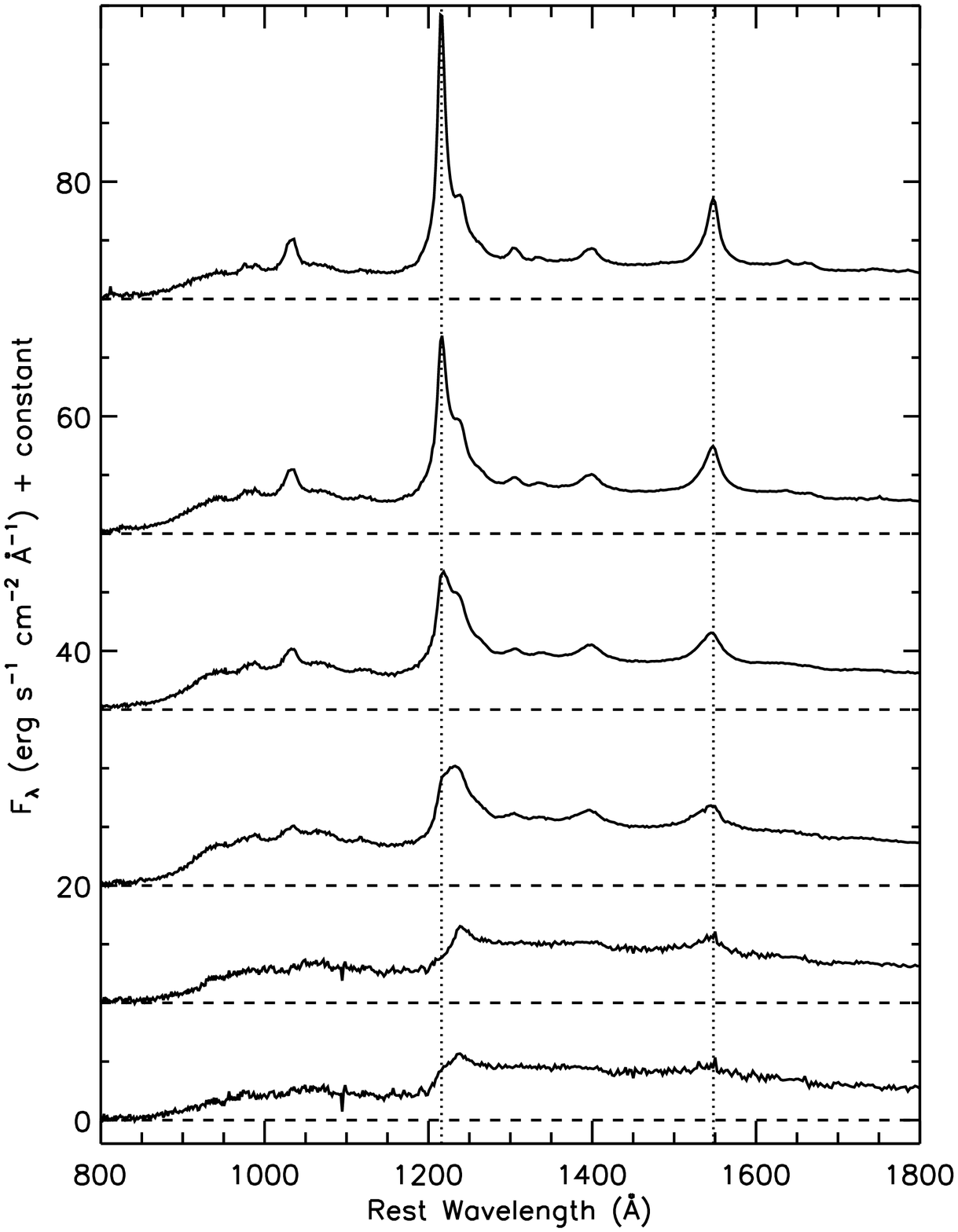}
\caption{Composite spectra of $z>3$ quasars in different EW(\lya+\nv)
bins.  The wavelengths of \lyalpha and \cfour are marked by dotted
lines.  Top panel: Composite for 428 quasars with EW $>1\sigma$ above
the mean ($\ew>101.9$~\AA).  Second panel: composite for 1038 quasars
with EW between the mean and $1\sigma$ above it
($63.6~\textnormal{\AA}<\ew<101.9$~\AA).  Third panel: composite for
1000 quasars with EW between the mean and $1\sigma$ below it
($39.6~\textnormal{\AA}<\ew<63.6$~\AA).  Fourth panel: composite for
536 quasars with EWs between $1\sigma$ and $3\sigma$ below the mean
($15.4~\textnormal{\AA}<\ew<39.6$~\AA).  Fifth panel: composite for 56
WLQs with EW $>3\sigma$ below the mean ($\ew<15.4$~\AA).  Bottom
panel: composite for the subset of 32 WLQs that do not exhibit
significant \cfour emission or PDLA absorption (see
Table~\ref{table:sample}).  There is a trend for the \lya/\nfive ratio
to decrease toward lower EWs.}
\label{fig:composite}
\end{center}
\end{figure}

\begin{figure}
\begin{center}
\includegraphics[angle=90,scale=0.35]{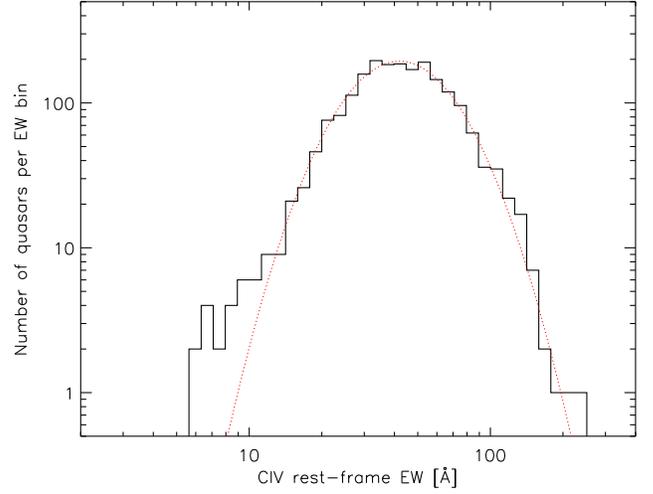}
\caption{Distribution of \cfour EWs for 2035/3058 quasars from
Figure~\ref{fig:ew_distribution} measured from emission-line fits
performed by \citet{shen08}.  The dotted red line shows the best-fit
log-normal distribution, with a mean EW of $41.9$~\AA\ and a $1\sigma$
range of $26.1$--$67.4$~\AA.  Objects with the weakest \cfour lines
are not included because accurate EWs could not be determined.}
\label{fig:civ_distribution}
\end{center}
\end{figure}

\section{The Weak Emission-Line Quasar Sample}\label{sec:sample}

The SDSS Fifth Data Release Quasar Catalog includes $77,429$ objects
and covers an area of 5740 deg$^2$ \citep{sch07}.  The 5374 quasars at
$z>3$\footnote{\lyalpha is detectable in SDSS spectra down to
$z\sim2.2$, but we focus our analysis on $z>3$ where the flux limit of
the quasar survey is a magnitude fainter \citep[e.g.,][see their
Figure 5]{ric06a}.  The highest-redshift quasar in the DR5 catalog is
at $z=5.41$.} are selected down to a magnitude limit of $i=20.2$
\citep{ric06a}.  The SDSS spectra have wavelength coverage
$\lambda=3800$--$9200$~\AA\ and resolution $R\sim2000$.  In
Figure~\ref{fig:example_spectra}, we show six spectra exhibiting a
range of redshifts, emission-line EWs, continuum spectral slopes, and
signal-to-noise ratios (S/Ns), including two WLQs.

For each spectrum, we fit a power law\footnote{The power-law spectral
slope of quasars is often expressed in terms of $\alpha$, where
$f_{\nu}\propto\nu^{\alpha}$.  We use this definition of $\alpha$
throughout the paper.  Here, $\beta=-\alpha-2$.} of the form
$f_{\lambda}=C\times\lambda^{\beta}$ to regions of the continuum that
are generally uncontaminated by emission lines (1285--1295,
1315--1325, 1340--1375, 1425--1470, 1680--1710, 1975--2050, and
2150--2250~\AA).  We then integrate the flux above the power-law
continuum between $\lrest=1160$~\AA\ and $\lrest=1290$~\AA\ to
determine emission-line flux and EW.  This region includes the
\lya~$\lambda1216$, \nv~$\lambda1240$, and \siii~$\lambda1263$
emission lines, so we measure the strength of a blended emission
feature that is dominated by the \lyalpha and \nfive components.  The
median S/N of the spectra integrated over this wavelength region is
$140$ for line$+$continuum and $95$ for the continuum.

The EW(\lya+\nv) measurements for all 5374 quasars are given in
Table~\ref{table:all}.  The typical statistical uncertainty of each
measurement is $\sim1$--2~\AA.  There is also a systematic uncertainty
in the placement of the continuum level, which depends on the
wavelength regions chosen for the continuum fit.  Visual inspections
show that the fits accurately describe the data, and while the EWs of
individual objects do change slightly ($\sim10\%$) if different
wavelength regions are used to fit the continuum, the aggregate
results are quite similar.  An important caveat is that our method
underestimates the intrinsic line strength when strong intervening
absorption is present.  Some studies \citep[e.g.,][]{die02}
interpolate across \lyalpha forest absorption features and fit the
intrinsic emission-line profile, but this requires an accurate
determination of the absorption systems along the line of sight to
each quasar and is difficult without high-S/N data at higher spectral
resolution.  The effects of \lyalpha forest absorption are discussed
further below and in Section~\ref{sec:redshift_distribution}.  In
addition, there is a break in the quasar continuum near
$\lrest=1100$~\AA\ \citep[e.g.,][]{zhe97, sha05}, so our power-law
continuum fit at $\lrest>1285$~\AA\ may slightly overestimate the
continuum flux at \lyalpha and therefore underestimate the strength of
the emission feature.

We restrict our analysis to the $3979$ sources that were identified as
primary quasar targets by the final algorithm given by \citet{ric02a}.
These objects have the uniform selection flag set in the \citet{sch07}
catalog and constitute a statistical color-selected sample.  We
additionally exclude 915 broad-absorption line (BAL) quasars from the
catalogs of \citet{tru06} and \citet{gib09}, as well as six objects
flagged by visual inspection as having either BAL-like absorption or
missing data in the \lyalpha region.  The final sample includes a
total of 3058 quasars.

The distribution of \lya+\nfive rest-frame EWs for this sample is
shown in Figure~\ref{fig:ew_distribution}.  The median and mean of the
measured values are 62.3~\AA\ and 67.6~\AA, respectively.  The
distribution is well described by a log-normal function with best-fit
parameters $\langle\log\ew\rangle=1.803$ and $\sigma(\log\ew)=0.205$,
corresponding to a mean of $63.6$~\AA\ and a $1\sigma$ range of
39.6--101.9~\AA.  However, there is a prominent tail toward small
values.  If the $\log$(EW) values were indeed distributed like a
Gaussian, where the $3\sigma$ range encompasses 99.73\% of the curve
area, we would expect to find $\sim4$ objects beyond this range
($\ew=15.4$--261.9~\AA) at each end.  Instead we find none at the high
end and 56 at the low end, and we define our sample of WLQs to be
these 56 objects with $\ew<15.4$~\AA.  This WLQ definition is more
inclusive than the $\ew<10$~\AA\ definition adopted by \citet{shem09},
but the precise cutoff between WLQs and ``normal'' quasars is somewhat
arbitrary; our goal is simply to select the sources with extreme EWs
that exist beyond what would be expected from a simple log-normal
distribution.

To illustrate the spectroscopic properties of WLQs relative to the
rest of the quasar sample, we construct composite spectra in bins of
EW(\lya+\nv) and present them in Figure~\ref{fig:composite}.  These
composite spectra span the range $\lrest=800$--1800~\AA\ and are
constructed following the method of \citet{fan04}; each spectrum is
normalized based on its flux at $\lrest=1450$~\AA\ and the average of
all spectra is computed.  We use five EW bins: (1) EW $>1\sigma$ above
the mean, 428 objects; (2) EW between the mean and $1\sigma$ above it,
1038 objects; (3) EW between the mean and $1\sigma$ below it, 1000
objects; (4) EW between $1\sigma$ and $3\sigma$ below the mean, 536
objects; and (5) EW $>3\sigma$ below the mean, 56 objects.  In the
first three composite spectra, the red side of the \lya+\nfive blend
(dominated by \nv) is nearly identical, but the blue side (dominated
by \lya) steadily decreases in strength; while in the fourth spectrum,
the peak flux shifts redward of $\lrest=1216$~\AA\ as the \nfive
component begins to dominate.  This trend continues in the WLQ
composite, where the 1216~\AA\ flux actually drops below the
(unabsorbed) continuum level.  This extreme behavior is driven by a
handful of WLQs with strong absorption features, and in fact 15/56
WLQs are included in the \citet{pro08} catalog of proximate damped
\lyalpha absorbers (PDLAs, $\Delta v<3000$~km~s$^{-1}$,
$N_H>2\times10^{20}$~cm$^{-2}$).  In the bottom panel, we exclude
these WLQs whose weak \lya+\nfive lines can be explained by absorption
rather than intrinsic weakness, as well as 10 additional WLQs that
exhibit $\ew(\civ)>10$~\AA\ (these sources are flagged in
Table~\ref{table:sample}).  The resulting WLQ composite spectrum has
flux roughly equal to the continuum level near \lya~$\lambda$1216, a
weak emission feature near \nv~$\lambda$1240, and a negligible bump
near \civ~$\lambda$1549.

There is also a trend for the continuum luminosity to increase in the
lower EW bins.  The median continuum luminosity for the whole sample
is 41\% larger than the median for sources in the highest EW bin and
39\% smaller than the median for WLQs.  This trend is consistent with
the Baldwin effect \citep[the observed anti-correlation between \cfour
EW and quasar luminosity;][]{bal77}.

We select the WLQ sample based on extreme \lya+\nfive EWs rather than
\cfour because the former line is stronger, and therefore readily
detected even at low S/N, and observable at the highest redshifts
($z>4.9$).  However, we do compile \cfour EWs measured from
emission-line fits performed by \citet{shen08}.  These values are
presented in Table~\ref{table:all}, and their distribution is shown in
Figure~\ref{fig:civ_distribution}.  For the 2035/3058 quasars in our
final sample where the \citet{shen08} fits yield secure \cfour EWs,
the best-fit log-normal distribution has parameters
$\langle\log\ew\rangle=1.622$ and $\sigma(\log\ew)=0.206$ (mean
41.9~\AA, $1\sigma$ range 26.1--67.4~\AA), and the median EW ratio
(\lya+\nv)/\cfour is 1.53.  This ratio is smaller than what has been
reported in previous studies
\citep[e.g.,][]{sch91,osm94,bro01,van01,die02}, which tend to find
larger EW(\lya+\nv) and smaller or comparable EW(\civ) values.  As
mentioned above, our method underestimates the intrinsic strength of
\lya+\nfive emission when strong \lyalpha forest absorption is
present, and the \citet{shen08} fits overestimate EW(\civ) when strong
\fetwo emission mimics a broad Gaussian component of the \cfour
profile; but our results for both lines are nonetheless within the
range of literature values \citep[e.g.,][]{fra91}.  Also, early
studies \citep[e.g.,][]{sch91,osm94} were biased toward more luminous
quasars, so their lower EW(\civ) values could be explained by the
Baldwin effect.  There is no information about objects with the
weakest \cfour lines in Table~\ref{table:all} or
Figure~\ref{fig:civ_distribution} because accurate EWs could not be
measured from the \citet{shen08} fits; in fact, 34/56 WLQs and all
objects with $\ew(\civ)<6$~\AA\ are not included.

\subsection{Sample Properties}\label{sec:wlq_results}

To investigate the nature of the low-EW tail of the distribution, in
this section we compare several properties of WLQs to those of the
general $z>3$ quasar population.  We perform this analysis on all 56
WLQs in the uniform sample and on the subset of 31 WLQs with no \cfour
or PDLA flags (see above and Table~\ref{table:sample}); the results
are insensitive to the choice of sample, and we quote values for the
full uniform sample.

\subsubsection{Emission-line and Continuum Properties}\label{sec:line_continuum}

In addition to EW, our continuum fits to the SDSS spectra yield values
of \lya+\nfive emission-line flux, continuum flux, and continuum
spectral slope for each source.  We find that the emission-line
luminosities of WLQs are all at the low end of the distribution for
normal\footnote{We use the term ``normal'' hereafter to refer to
non-WLQs (i.e., quasars with $\ew>15.4$~\AA).} $z>3$ quasars (28/56
WLQs are in the bottom 1\%, 44/56 are in the bottom 5\%, and all are
in the bottom 26\%); by the Kolmogorov--Smirnov (K-S) test, the
probability that the two samples are drawn from the same parent
distribution is $p=4\times10^{-31}$.  The median line luminosity for
WLQs is a factor of $3.8$ lower than the median for normal quasars,
indicating that \lyalpha is weak in an absolute sense, rather than
just relative to the continuum.  As mentioned above, the continuum
luminosities for WLQs are somewhat larger on average than those for
normal quasars, but their continuum slopes are quite similar.  We find
median values of $\alpha=-0.54$ and $\alpha=-0.52$ for normal quasars
and WLQs, respectively, and the K-S test gives a probability $p=0.20$
that the distributions are the same.

\subsubsection{Radio Flux}\label{sec:radio_loudness}

We also investigate the radio properties of WLQs using data from the
Faint Images of the Radio Sky at Tweny cm (FIRST) survey
\citep{bec95}.  A search of the catalog \citep{whi97} within
$2\arcsec$ of each quasar yields 147 detections and 2809 upper
limits\footnote{We calculate the detection threshold corresponding to
the rms at each position.  This is a $5\sigma$ upper limit, including
CLEAN bias: $F_{\nu}=5\sigma+0.25$~mJy \citep{whi97}.}; 102 quasars,
including four WLQs, are not covered by the FIRST footprint.  We find
that the radio-detection fraction is $4.9$\% (142/2904) for normal
quasars and 10\% (5/52) for WLQs.  This difference is not significant;
given the radio-detection fraction of normal quasars, the binomial
distribution gives a $11\%$ chance of having five or more detections
in a sample of 52.  If we relax the uniform selection criterion
described above and include all non-BAL quasars, the radio fraction
increases to 6.1\% (233/3811) for the general population and to $21$\%
(15/70) for WLQs.  This increase, which is especially prominent for
WLQs, is due to the fact that FIRST sources were targeted for
spectroscopy by the SDSS even if their optical colors did not meet the
final quasar-selection criteria.  Assuming a 6.1\% detection
probability, the chance of having 15 or more detections in a sample of
70 is only $0.002$\%.

A more relevant quantity than the radio detection fraction is the
radio-loudness parameter $R$, which is defined as the ratio of radio
to optical flux density.  We adopt the definition\footnote{The
original definition of $R$ by \citet{kel89} measures the optical flux
density at the $B$ band, $\lambda=4400$~\AA.  We choose a shorter
wavelength to more closely match the rest-frame wavelengths observed
by the SDSS.  For a power-law slope $\alpha=-0.5$, one would expect
$33$\% more flux density at 4400~\AA\ than at 2500~\AA\, so the
definition we use generally produces larger $R$ values.} used by
\citet{jia07}, $R=f_{\nu}(6~$cm$)/f_{\nu}(2500~$\AA$)$.  We calculate
the rest-frame 6~cm flux density from the observed 20~cm flux density
assuming a power-law slope $\alpha_r=-0.8$, and we calculate the
rest-frame 2500~\AA\ flux density from the observed $z$ band
($\lambda=8931$~\AA) flux density assuming a power-law slope
$\alpha_o=-0.5$.  We find that 3.2\% (92/2904) of normal $z>3$ quasars
and 6\% (3/52) of WLQs have $R>100$, while $>4.6$\% (133/2904) and
$>10$\% (5/52) have $R>10$.  The latter fractions are lower limits
because the FIRST detection threshold, typically $\sim1$~mJy, is only
deep enough to constrain $R<10$ for the optically brightest sources
($\sim15$\% of the sample).  Again, including all non-BAL quasars
changes the $R>10$ fraction to $>5.9$\% (224/3811) and $>21$\%
(15/70), and the $R>100$ fraction to 4.1\% (156/3811) and 7\% (5/70).

To compare the distributions of $R$ values statistically, given that
most values are upper limits (i.e., the distributions are heavily
censored), we use the survival analysis software package ASURV Rev 1.3
\citep{lav92}, which implements the methods presented in
\citet{fei85}.  We perform the Gehan, logrank, and Peto-Prentice
two-sample tests.  In the uniformly selected sample, these tests give
probabilities $p=0.097$, $p=0.145$, and $p=0.147$ that the WLQs and
normal $z>3$ quasars have the same distribution of radio loudness.  If
the uniform-selection criterion is dropped (i.e., all non-BAL quasars
are included), these probabilities drop to $p=1\times10^{-4}$,
$p=2\times10^{-6}$, and $p=8\times10^{-6}$.  Thus, there is no
evidence of a difference in radio properties between WLQs and normal
quasars among the sources that were selected for spectroscopy solely
on the basis of optical colors.  However, the use of FIRST data to
identify SDSS spectroscopic targets selects a larger fraction of WLQs;
of the non-BAL quasars at $z>3$ with FIRST detections, $6.0$\%
(15/248) have $\ew(\lya+\nv)<15.4$~\AA.  This may be because radio
selection is not biased against quasars with weak \lyalpha emission,
while color selection at $z<4$ is biased (see
Section~\ref{sec:redshift_distribution}).

\begin{figure}
\begin{center}
\includegraphics[angle=90,scale=0.35]{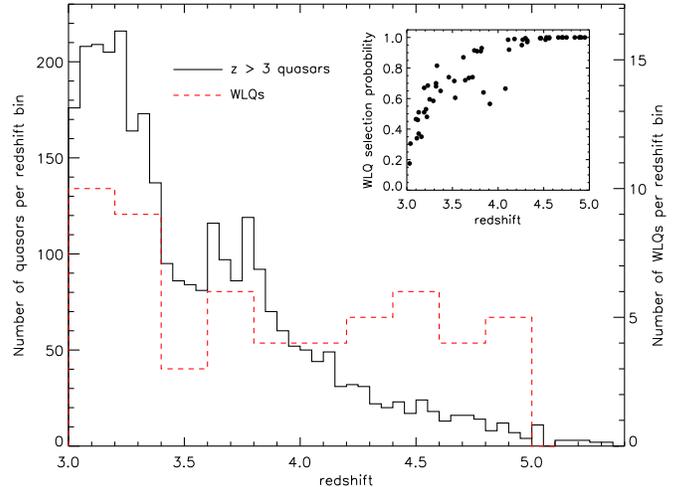}
\caption{Redshift distributions of normal $z>3$ quasars (solid black
line, bin size $\Delta z=0.05$) and WLQs (dashed red line, bin size
$\Delta z=0.20$).  The two distributions diverge at $z\sim4.2$, with
WLQs appearing to be more common at higher redshift.  The inset panel
shows the selection probability for each WLQ (see
Section~\ref{sec:redshift_distribution}), indicating that weak-lined
sources are more likely to be selected at higher redshift.  The
increased opacity of the IGM near $\lrest=1216$~\AA\ at higher
redshift is also likely to bias EW measurements toward lower values.}
\label{fig:redshift_distribution}
\end{center}
\end{figure}

\subsubsection{Redshift Distribution}\label{sec:redshift_distribution}

The redshift distributions of normal $z>3$ quasars and WLQs are shown
in Figure~\ref{fig:redshift_distribution}.  The two distributions
diverge at $z\sim4.2$, with WLQs constituting $1.3$\% ($36/2737$) of
the quasar population below this redshift and $6.2$\% ($20/321$) above
it.  The redshifts measured for individual WLQs can be uncertain by as
much as $\delta z=0.1$ because of their weak lines (see
Section~\ref{sec:optical_analysis}), but this does not affect the
result that WLQs are a larger fraction of the observed quasar
population at higher redshift.  The typical \lya+\nfive EW for normal
quasars at $z>4.2$ is also lower than for lower redshift quasars, with
median values $46.4$~\AA\ and $63.8$~\AA, respectively.  The greater
incidence of lower EW objects at higher redshift could be caused
either by selection effects or by some differential physical process
(see below).  To investigate the former hypothesis, we calculate the
selection probability of each WLQ based on its redshift, $i$-band
luminosity, spectral slope, and \lyalpha emission-line EW using the
method of \citet{fan01} and \citet{ric06a}.  Using a distribution
function of \hone absorption, we simulate 200 spectra, compute $ugriz$
magnitudes including photometric errors, and apply the
quasar-selection criteria.  The results are shown in the inset of
Figure~\ref{fig:redshift_distribution}.  We find that the selection
probabilities of individual WLQs generally increase with redshift,
with average probabilities of $p=0.53$ at $z<3.5$ and $p=1.0$ at
$z>4.5$.  This redshift-dependent selection can be explained by the
fact that absorption by the \lyalpha forest and Lyman Limit Systems
(LLSs) equalizes the colors of higher-redshift quasars, independent of
their emission line properties (see Section~\ref{sec:intro}).  There
is also significant dependence on the luminosity and spectral slope,
as well as emission-line EW, at each redshift \citep[e.g.,][see their
Figure 5]{fan01}, but the general result is that WLQs are easier to
select at higher redshift.  The fact that the observed WLQ fraction is
$\sim6$\% in samples that are unbiased with respect to emission-line
strength (e.g., FIRST-detected quasars, quasars at $z>4.2$) implies
that the intrinsic WLQ fraction may be significantly higher than the
$\sim1$\% inferred from the full SDSS sample.

In addition to affecting the colors of WLQs, and thus their likelihood
of selection, absorption by the \lyalpha forest can also affect the
\lyalpha emission line itself, pushing objects toward lower EWs at
higher redshift.  To have a measurable effect, the absorbing material
must be located in close proximity to the quasar,\footnote{We measure
\lyalpha flux at $\lrest>1160$~\AA, so absorbers must be within
$\Delta z=0.046$, corresponding to a peculiar velocity $\Delta
v=3100$~km~s$^{-1}$ or a physical distance $D\sim8$~Mpc at $z=3.5$,
and $\Delta v=2500$~km~s$^{-1}$ or $D\sim5$~Mpc at $z=4.5$.} as is
clearly the case for 11 WLQs that have PDLAs.  However, there is no
evidence to suggest that PDLAs, which are generally associated with
galaxies in the environment of the quasar \citep[e.g.,][]{mol98,
ell02, rus06}, are more common at higher redshift, and an analysis of
the evolution of the intergalactic medium (IGM) opacity at $z>3$ is
beyond the scope of this paper.  The relevant effect of the \lyalpha
forest effect on the \lyalpha emission line, though, can be seen in
the composite spectrum of $z\sim6$ quasars \citep{fan04}, and it is
likely to bias the EW measurements of the highest-redshift sources.

\begin{figure*}
\begin{center}
\includegraphics[angle=90,scale=0.25]{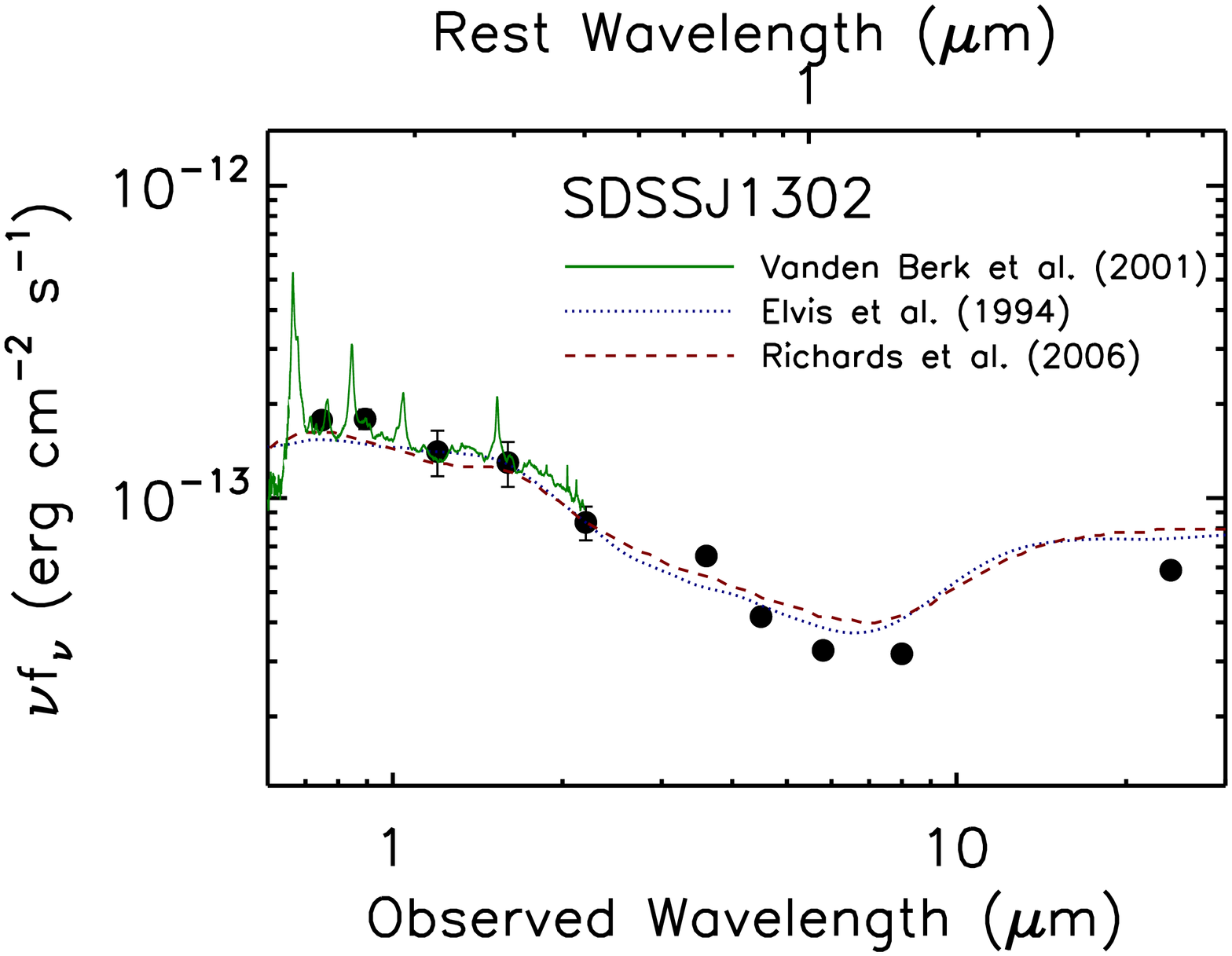}
\includegraphics[angle=90,scale=0.25]{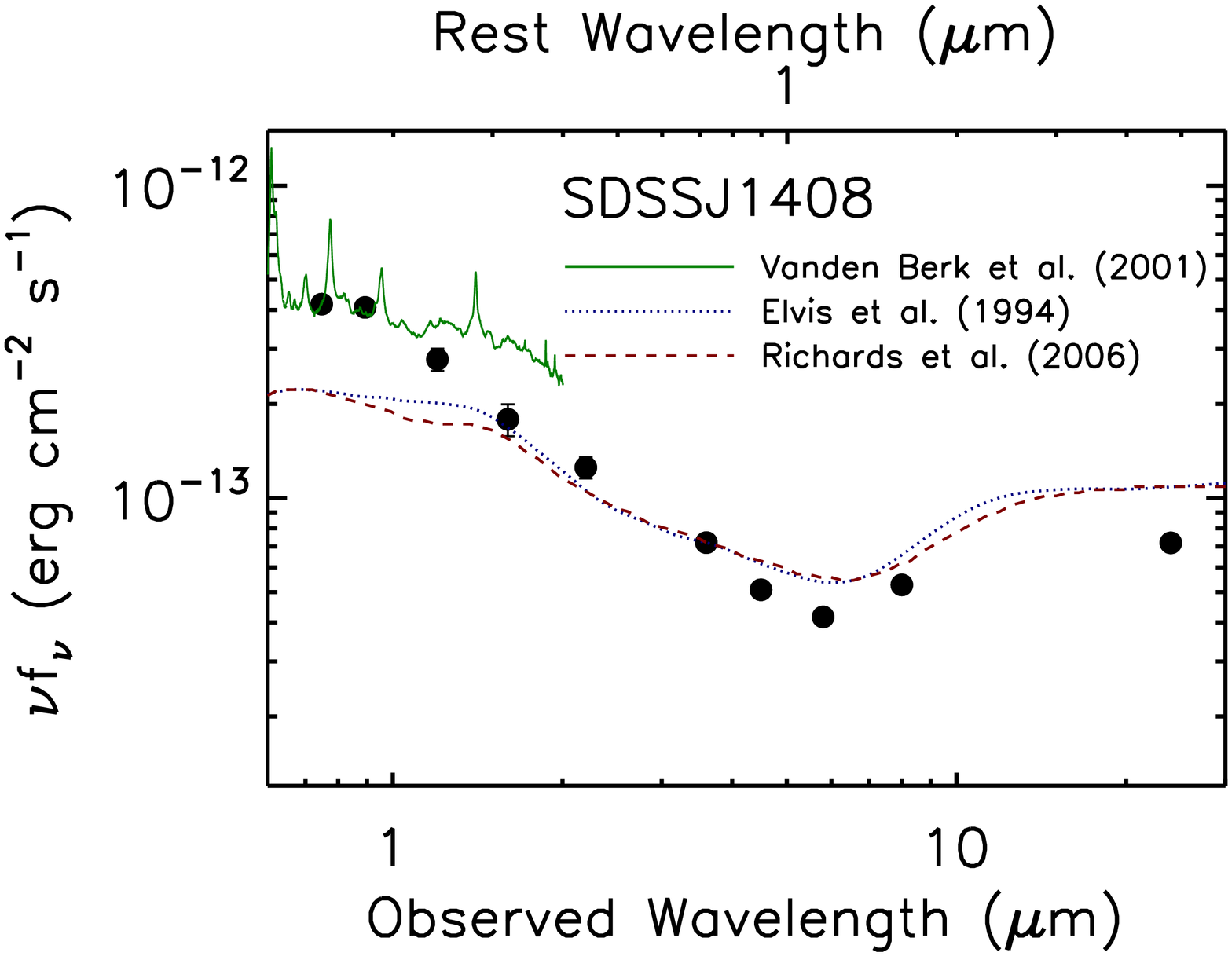}
\end{center}
\begin{center}
\includegraphics[angle=90,scale=0.25]{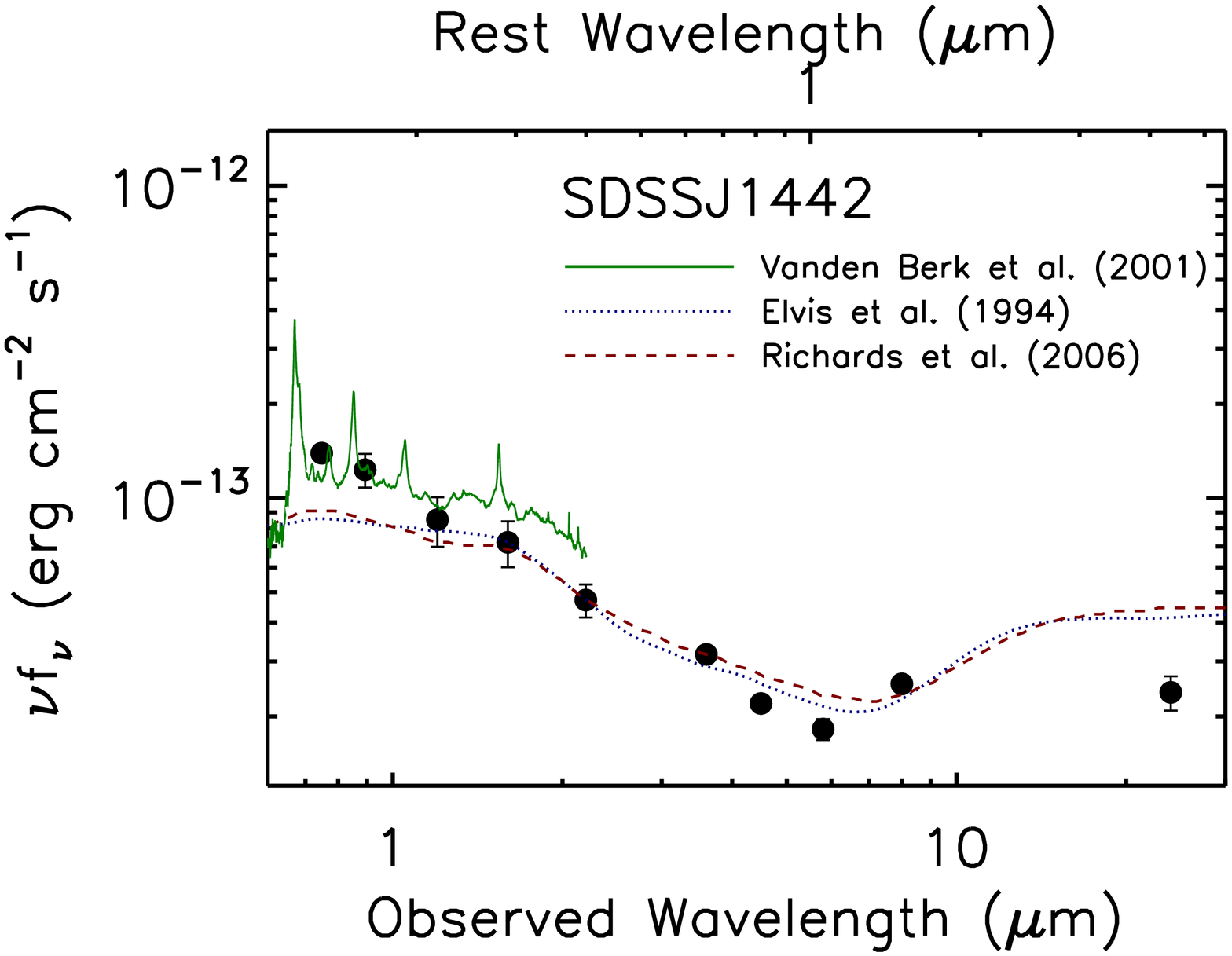}
\includegraphics[angle=90,scale=0.25]{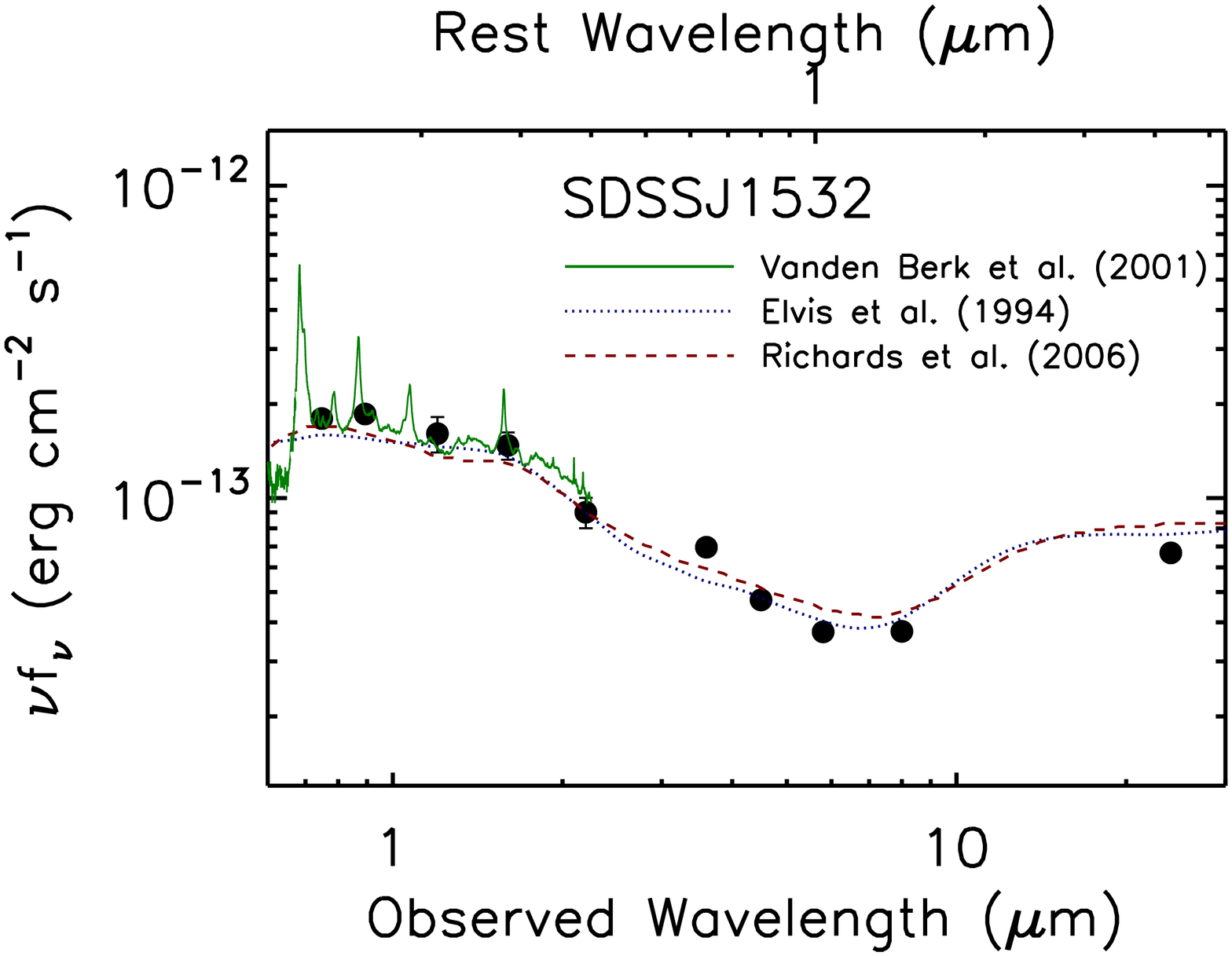}
\caption{Observed-frame $0.7-24~\mu$m SEDs for four representative
  WLQs at $z>4$.  The data points are SDSS $i$ and $z$ bands; $J$,
  $H$, and $K_{s}$ bands; {\it Spitzer} IRAC 3.6, 4.5, 5.8, and
  8.0~$\mu$m; and {\it Spitzer} MIPS 24~$\mu$m.  For most data points,
  the error bars are smaller than the dots.  Each panel includes the
  \citet{van01} quasar composite spectrum scaled to the SDSS data
  points (green solid line), and the mean quasar composite spectra
  from \citet{elv94} and \citet{ric06b} scaled to all of the data
  points (blue dotted and red dashed lines, respectively).}
\label{fig:seds_template}
\end{center}
\end{figure*}

\begin{figure*}
\begin{center}
\includegraphics[angle=90,scale=0.25]{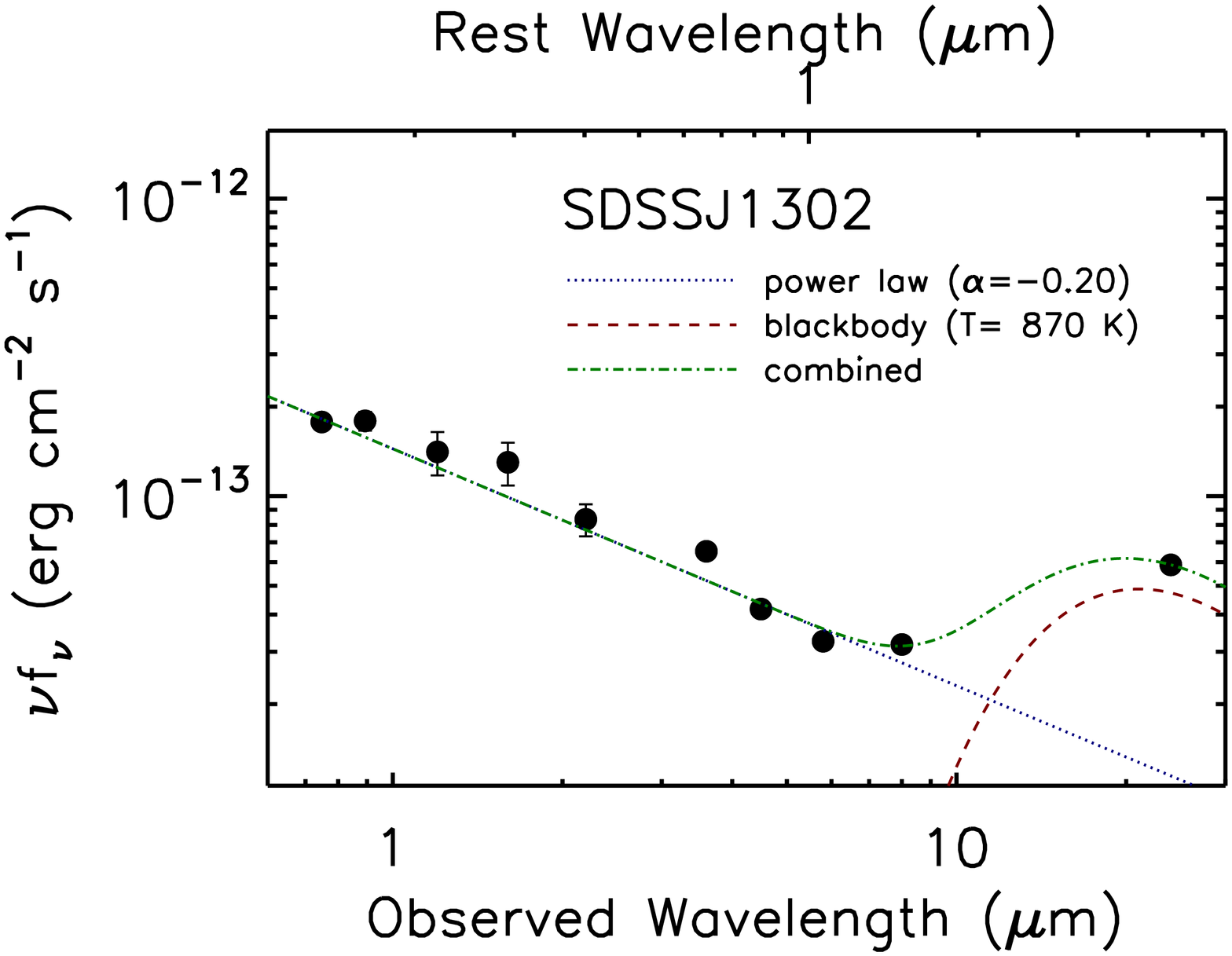}
\includegraphics[angle=90,scale=0.25]{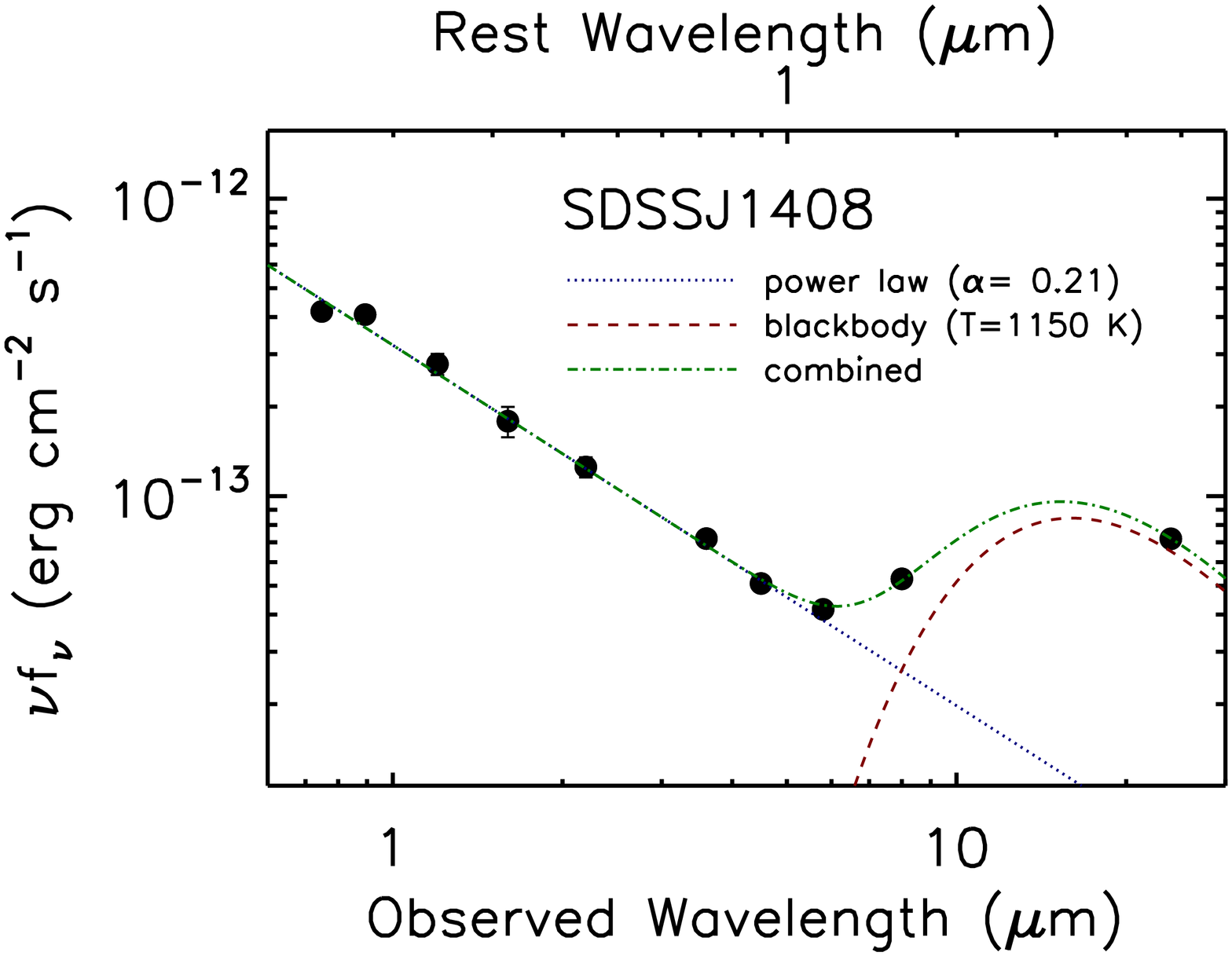}
\end{center}
\begin{center}
\includegraphics[angle=90,scale=0.25]{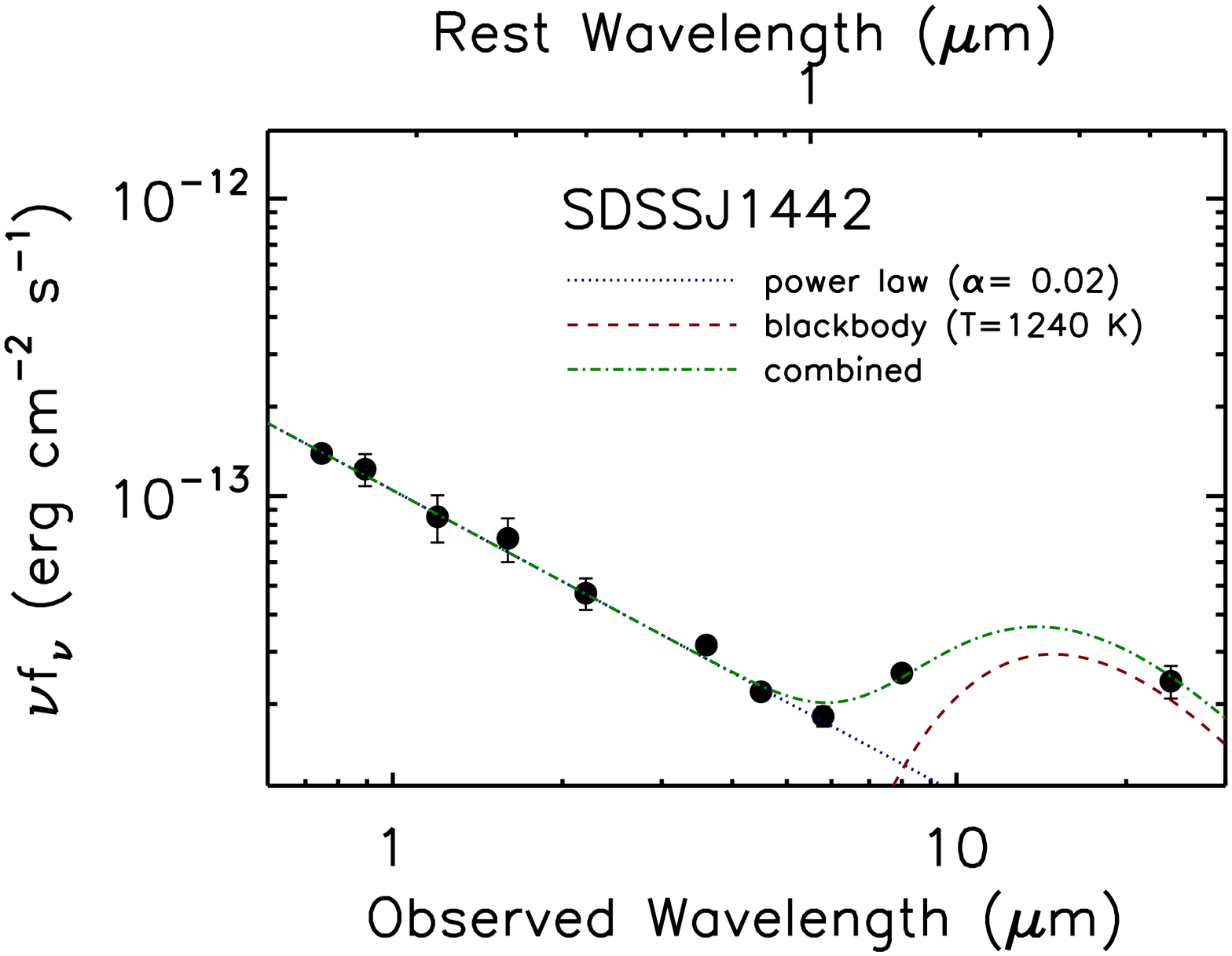}
\includegraphics[angle=90,scale=0.25]{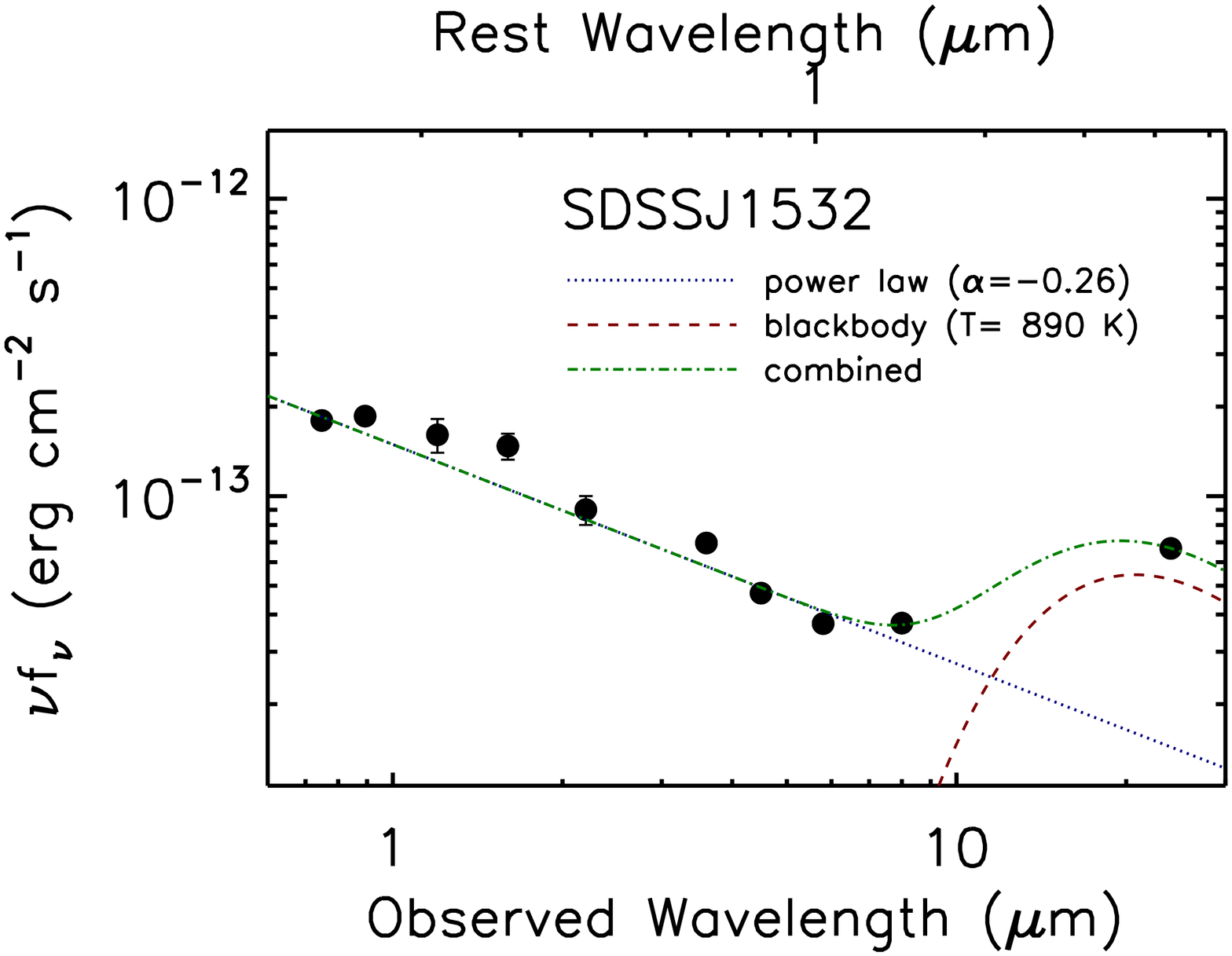}
\caption{Data are the same as Figure~\ref{fig:seds_template}.  The
  dotted blue line shows the best-fit power-law model to the
  short-wavelength data and the dashed red line shows the best-fit
  single-temperature blackbody to the longer wavelength data.  The
  dash-dotted green curve is the sum of the two models.  Also included
  are the spectral slope of the power law and the temperature of the
  blackbody for each source.}
\label{fig:seds_fit}
\end{center}
\end{figure*}

\section{IR Photometry}\label{sec:ir}

This is the first of three sections (Sections~\ref{sec:ir},
\ref{sec:optical}, \ref{sec:synchrotron}) that present multiwavelength
observations of four $z>4$ WLQs: SDSSJ1302, SDSSJ1408, SDSSJ1442, and
SDSSJ1532.  These sources were selected early on in the survey to be
representative of the handful of WLQs known at the time; their
\lyalpha emission-line strengths range from undetectable (e.g.,
SDSSJ1532) to weak, but clearly detected (e.g., SDSSJ1442), and two
are FIRST radio sources (SDSSJ1408, $R=14$; SDSSJ1442, $R=26$), while
the other two have no radio detections.  The FIRST-detected sources
also have strong {\it Chandra} X-ray detections \citep[SDSSJ1408,
$\alpha_{ox}=-1.54$; SDSSJ1442, $\alpha_{ox}=-1.42$][]{shem06}, while
SDSSJ1302 and SDSSJ1532 are X-ray weak \citep[$\alpha_{ox}=-2.08$ and
$\alpha_{ox}<-1.90$, respectively;][]{shem06,shem09}.  This section
presents {\it Spitzer} mid-IR photometry in the 3.6--24~$\mu$m range
(Section~\ref{sec:spitzer}) and near-IR photometry in $J$-, $H$-, and
$K_{s}$-band (Section~\ref{sec:nir}), which we use to construct
rest-frame 0.1--5~$\mu$m SEDs of WLQs that can be compared to those of
normal quasars (Section~\ref{sec:seds}).

\subsection{{\it Spitzer} Mid-IR Data}\label{sec:spitzer}

We obtained mid-IR photometry for all four sources as part of a {\it
Spitzer Space Telescope} \citep{wer04} Cycle I General Observer
Program (PID 3221).  Observations with the IRAC instrument
\citep{faz04} were executed in all four channels ($3.6~\mu$m,
$4.5~\mu$m, $5.8~\mu$m, and $8.0~\mu$m) with an integration time of
1000~s per channel.  Observations with MIPS \citep{rie04} were
obtained in the $24~\mu$m channel with integration times of
1400--1500~s.

We begin our analysis of the IRAC data using the Basic Calibrated Data
(BCD) products from version S14.0.0 of the {\it Spitzer} Science
Center software pipeline.  The MOPEX software package was used to
construct mosaics from the individual frames, which involves
background matching between overlapping images and outlier pixel
rejection.  The native IRAC pixel size of $1.2\arcsec$ yields somewhat
undersampled data, so we take advantage of a subpixel dither pattern
to interpolate the individual BCD images onto a common grid with a
$0.6\arcsec$ pixel scale using the Drizzle algorithm.  We perform
aperture photometry on the mosaics using the APER task in IDLPHOT.  We
use $3.6\arcsec$ source apertures with the sky sampled in a
14.4--20$\arcsec$ annulus.  Aperture corrections are calculated based
on the IRAC point-spread function (PSF) and applied to the flux
density measurements.  The statistical uncertainties in our
measurements are calculated based on Poisson noise in the source
counts, uncertainty in the background level, and noise from sky
variations in the source aperture.  Except for SDSSJ1442 in channel 3
and channel 4, the statistical uncertainties for all mosaics are
smaller than the 5\% calibration accuracy of IRAC \citep{rea05}, so we
quote a 5\% uncertainty on all of our flux density values.  We also
calculate array-location-dependent photometric corrections for each
mosaic; the differences between photometry performed on the corrected
and uncorrected mosaics are quite small in channels 1 and 2
(1\%--2~\%) and somewhat larger in channels 3 and 4 (1\%--5~\%).  We
quote the average source flux density from the corrected and
uncorrected mosaics in Table~\ref{table:spitzer_photometry}, which is
appropriate for sources that are flat in $f_{\nu}$ through the IRAC
bands.  We do not apply any other color corrections or pixel phase
corrections because we found them to be $<1$~\% effects.

We are unable to perform accurate aperture photometry on the IRAC data
for SDSSJ1532 because of the presence of a star $3.7\arcsec$ away
($3.4\arcsec$ east, $1.4\arcsec$ south).  Based on the available
near-IR (Section~\ref{sec:nir}) and {\it Spitzer} data, this source
appears to be an M star that peaks at the $K_{s}$ band
($f_{\nu}\sim80~\mu$Jy) and is fainter at longer wavelengths
($f_{\nu}\sim70~\mu$Jy at IRAC channel 1, $f_{\nu}\sim45~\mu$Jy at
channel 2).  We use the APEX software package to perform point-source
fitting photometry on the SDSSJ1532 mosaic images (accurate to 5\%),
and present these results in Table~\ref{table:spitzer_photometry}.

Our analysis of the MIPS data is based upon the BCD products from
version S14.4.0 of the software pipeline.  We discard the first three
data collection events (DCENUM=0,1,2) for each Astronomical
Observation Request to alleviate the first-frame effect.  We perform
additional flat-fielding with {\it flatfield.pl} in MOPEX to remove a
background gradient across the frames.  We then create mosaics,
matching backgrounds and removing outlier pixels.  We perform aperture
photometry on each mosaic using a $6\arcsec$ source radius and a
20--32$\arcsec$ sky annulus.  Aperture corrections are applied to the
measured flux densities based on the MIPS $24~\mu$m PSF.  We also
perform point-source fitting photometry on all the mosaics, and find
results consistent with the aperture photometry, within the
uncertainties.  We apply a $4$\% color correction (i.e., we divided
the flux density measurements by $0.96$), which is appropriate for a
$f_{\nu}\sim\nu^{-1}$ power law typical for quasars in this rest-frame
frequency range \citep{elv94,ric06b}.  Our results are presented in
Table~\ref{table:spitzer_photometry}.

\subsection{Ground-based Near-IR Data}\label{sec:nir}

We obtained $J$-band photometry for all four targets and $H$-band
photometry for SDSSJ1302 and SDSSJ1408 on 2005 April 26--27 with the
$256\times256$ near-IR camera on the Steward Observatory Bok 2.3-m
telescope.  The camera uses a Near-Infrared Camera and Multi-Object
Spectrometer (NICMOS) array \citep{rie93} and has a plate scale of
$0.6\arcsec$ pix$^{-1}$.  The data were taken in nine-position
dithered sequences of 60-s exposures.  The seeing ranged from
1.3$\arcsec$ to 2.0$\arcsec$, and we use 1.8--3.6$\arcsec$ source
apertures and a 6--12$\arcsec$ sky annulus for photometry.

We acquired $K_s$-band photometry for SDSSJ1302 on 2005 April 30 and
for SDSSJ1408 and SDSSJ1442 on 2005 May 1 with PANIC \citep{mar04} on
the 6.5-m Baade telescope at Las Campanas Observatory.  PANIC utilizes
a Rockwell $1024\times1024$ IR Hawaii detector with a plate scale of
$0.125\arcsec$ pix$^{-1}$.  The data were taken in five-position
dithered sequences of 20-s exposures.  The seeing ranged from
0.3$\arcsec$ to 0.5$\arcsec$, and we use 0.6--1.9$\arcsec$ source
apertures and a 2.5--3.1$\arcsec$ sky annulus for photometry.

All four objects were observed in $H$ and $K_{s}$ band, and all but
SDSSJ1532 were observed in $J$ band on 2005 June 30 using the Near-IR
Camera/Fabry-Perot Spectrometer \citep[NIC-FPS,][]{hea04} on the
Apache Point 3.5-m telescope.  The camera has a Rockwell Hawaii-IRG
$1024\times1024$ CCD with a plate scale of 0.273$\arcsec$~pix$^{-1}$.
The data were taken in five-position dithered sequences in a
20$\arcsec$ box surrounding the center of the chip, with 60-s
exposures in the $J$ band, 20-s exposures in the $H$ band, and 10-s
exposures in the $K_{s}$ band.  The seeing during the observations of
SDSSJ1302 and SDSSJ1408 ranged from 0.7$\arcsec$ to 0.9$\arcsec$, and
the seeing during the observations of SDSSJ1442 and SDSSJ1532 ranged
from 1.3 to 1.5$\arcsec$.  We use 1.1--2.7$\arcsec$ source apertures
with a 2.7--5.5$\arcsec$ sky annulus for data obtained in good seeing
and a 5.5--6.8$\arcsec$ annulus for data obtained in poorer seeing.

For all of the near-IR data, we stack and sky-subtract dithered
sequences of exposures using procedures included in the PANIC IRAF
package.  For each sequence, we form a median sky frame that is scaled
to each image and subtracted.  We perform aperture photometry on the
science targets and calibrate the measurements using sources in the
field from the Two Micron All Sky Survey (2MASS) catalog \citep{skr06},
and standard stars from the catalog of \citet{per98} that were
observed when conditions were photometric.  The uncertainties on these
flux density measurements are 5\%--20\%.  We quote the photometry from
each epoch and combine all results into a weighted average flux
density for each source in each band in
Table~\ref{table:nir_photometry}.  Our science targets are too faint
to be detected by 2MASS, so these are the only photometric epochs
available.

\subsection{Rest-frame 0.1--5~$\mu$m SEDs}\label{sec:seds}

The observed SEDs from SDSS $i$ band through MIPS $24~\mu$m are shown
in Figures~\ref{fig:seds_template} and \ref{fig:seds_fit}.
Figure~\ref{fig:seds_template} also shows the SDSS quasar composite
spectrum from \citet{van01} and the mean quasar SEDs from
\citet{elv94} and \citet{ric06b}.  The data points are from different
epochs, so variability may be an issue, but any fluctuations are
likely to be $<10$\% (see Section~\ref{sec:optical_analysis}).  We
normalize the SDSS composite to match the $i$- and $z$-band points,
and we normalize the template SEDs to provide an overall match to all
of the data points.  The normalized templates are reasonably good
descriptions of the data for the two radio-undetected objects,
SDSSJ1302 and SDSSJ1532, in the sense that no data point deviates from
the template by more than 30\% over the full range of wavelengths from
$\lrest=1300~$\AA\ to $\lrest=5~\mu$m.  The two radio-detected
objects, SDSSJ1408 and SDSSJ1442, are brighter at shorter wavelengths
by 90\% and 60\% at the $i$ band, and fainter at longer wavelengths by
30\% and 40\% at the MIPS band, respectively, than the mean quasar
SEDs, but given the $\sim0.5$ dex $1\sigma$ scatter of individual
quasars around the mean SEDs \citep{elv94, ric06a}, these are not
significant deviations.

In Figure~\ref{fig:seds_fit}, we fit the data points with a simple
model that includes a power-law fit to the short-wavelength
($\lrest<1~\mu$m) flux from the accretion disk (or a relativistic
jet), combined with a single-temperature blackbody fit to the longer
wavelength thermal emission from hot dust.  We allow the power-law
slope and the blackbody temperature to vary, and we fit both
simultaneously.  We find power-law slopes in the range
$-0.26<\alpha<0.21$ and dust temperatures in the range
$870$~K~$<T<1240$~K.  This model fits the SEDs of the radio-detected
quasars better ($\chi^2/\textnormal{dof}=1.7$ for SDSSJ1408 and 1.3
for SDSSJ1442) than it does the SEDs of the radio-undetected quasars
($\chi^2/\textnormal{dof}=4.7$ for SDSSJ1302 and 5.6 for SDSSJ1532).
The radio-detected sources are also both bluer and have higher
temperature dust, which is consistent with their redder $[5.8]-[8.0]$
colors.

\begin{figure}
\begin{center}
\includegraphics[angle=0,scale=0.4]{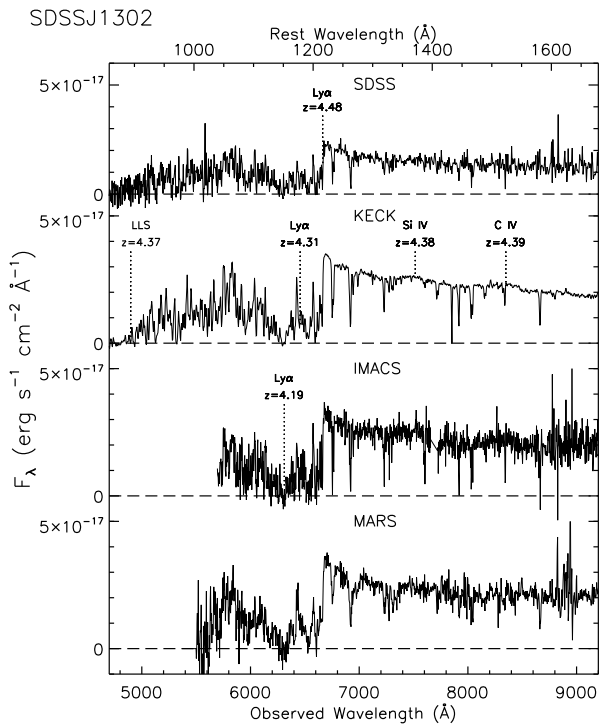}
\caption{Four epochs of optical spectroscopy for SDSSJ1302.  The SDSS
  spectrum ($R\sim2000$) is binned by a factor of 3, the IMACS
  spectrum ($R\sim3000$) is binned by a factor of 5, and the MARS
  spectrum ($R\sim750$) is unbinned.  Assuming the break associated
  with the onset of the \lyalpha forest is at $\lrest=1216$~\AA, we
  measure $z=4.48$ as the systemic redshift.  In the second panel, we
  identify an LLS at $z=4.37$ with associated \lyalpha absorption, as
  well as weak \sifour and \cfour emission lines that are blueshifted
  by $\sim5000$~km~s$^{-1}$ with respect to $z=4.48$.  In the third
  panel, we mark the \lyalpha component of a strong absorption-line
  system at $z=4.19$, which has metal lines \cii~$\lambda1334$,
  \siiv~$\lambda\lambda1393,1402$, \siii~$\lambda1526$, and
  \civ~$\lambda\lambda1548,1550$.  There are also two weaker
  absorption-line systems at $z=4.27$ and $z=3.98$ \lya, \siiv, and
  \civ.}
\label{fig:j1302}
\end{center}
\end{figure}

\begin{figure}
\begin{center}
\includegraphics[angle=0,scale=0.4]{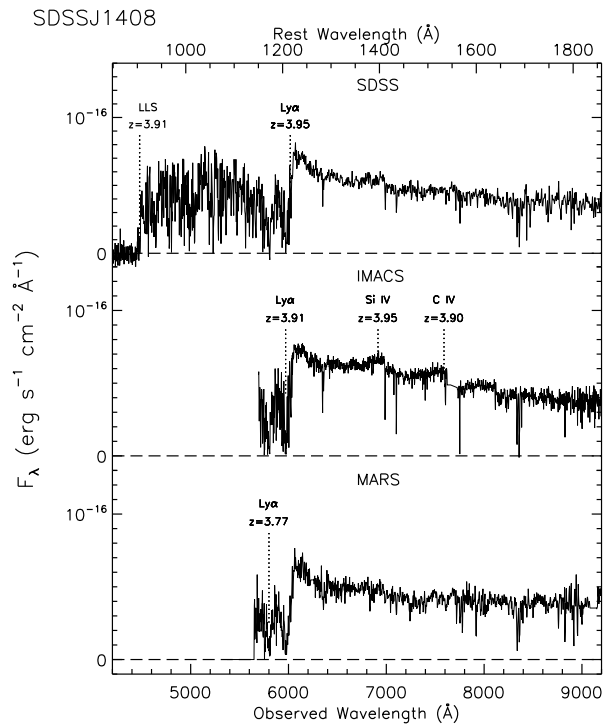}
\caption{Three epochs of optical spectroscopy for SDSSJ1408.  The
resolution and binning are as in Figure~\ref{fig:j1302}.  We measure
the systemic redshift to be $z=3.95$, which is lower than the redshift
given by SDSS $z=4.01$.  We identify an LLS at $z=3.91$ with
associated \lyalpha absorption.  Weak \sifour and \cfour emission
lines are detected in the second panel; the latter line may be
blueshifted by $\sim3000$~km~s$^{-1}$, but the gap between the IMACS
chips make this highly uncertain.  In the bottom panel, we mark a
\lyalpha absorber at $z=3.77$.}
\label{fig:j1408}
\end{center}
\end{figure}

\begin{figure}
\begin{center}
\includegraphics[angle=0,scale=0.4]{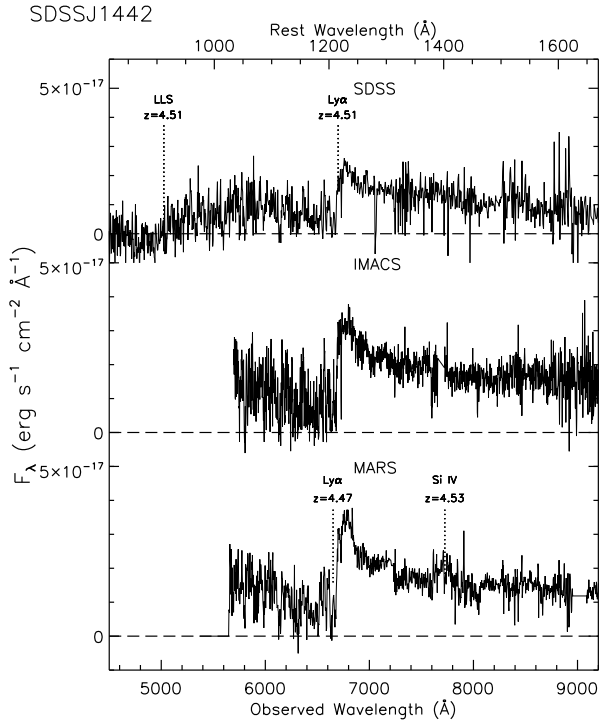}
\caption{Three epochs of optical spectroscopy for SDSSJ1442.  The
resolution and binning are as in Figure~\ref{fig:j1302}.  We measure
the systemic redshift to be $z=4.51$, and we identify an LLS at the same 
redshift.  The rest-frame \lyalpha EW increases with time from 
9~\AA\ in the SDSS spectrum to 16~\AA\ in the MARS spectrum.  We also detect 
\sifour emission in this latter epoch.  In the bottom panel, we mark a
\lyalpha absorber at $z=4.47$.}
\label{fig:j1442}
\end{center}
\end{figure}

\begin{figure}
\begin{center}
\includegraphics[angle=0,scale=0.4]{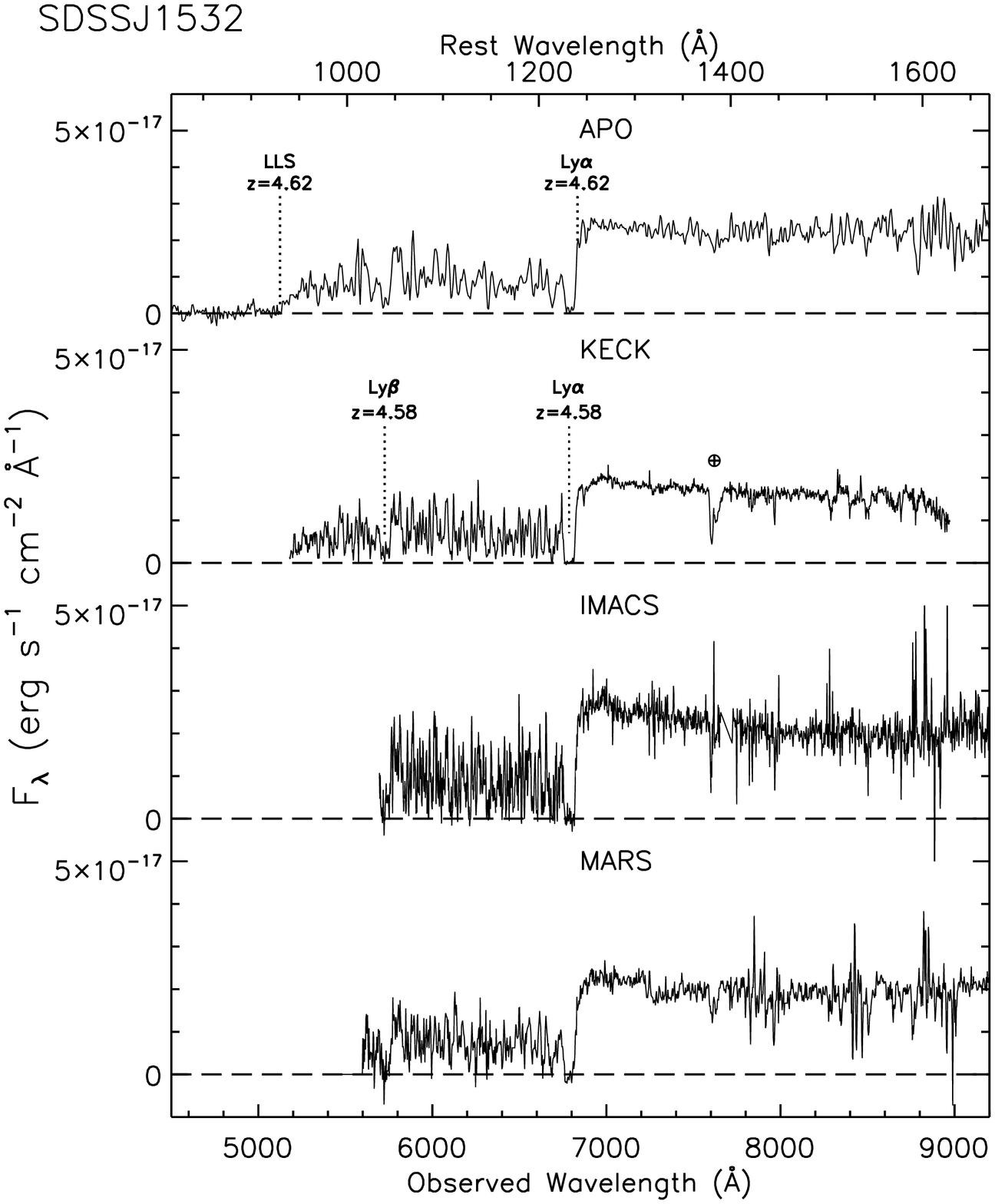}
\caption{Four epochs of optical spectroscopy for SDSSJ1532.  The
spectra in the top two panels are taken directly from \citet{fan99},
and the resolution and binning of the IMACS and MARS spectra are as in
Figure~\ref{fig:j1302}.  A redshift of $z=4.62$ is consistent with
both the onset of the \lyalpha forest and the location of the LLS.  In
the second panel, we mark a $z=4.58$ system with \lyalpha and
Ly$\beta$ absorption.  The rest-frame EW of \lyalpha is less than
3~\AA\ in all epochs. }
\label{fig:j1532}
\end{center}
\end{figure}

\begin{figure*}
\begin{center}
\includegraphics[angle=90,scale=0.25]{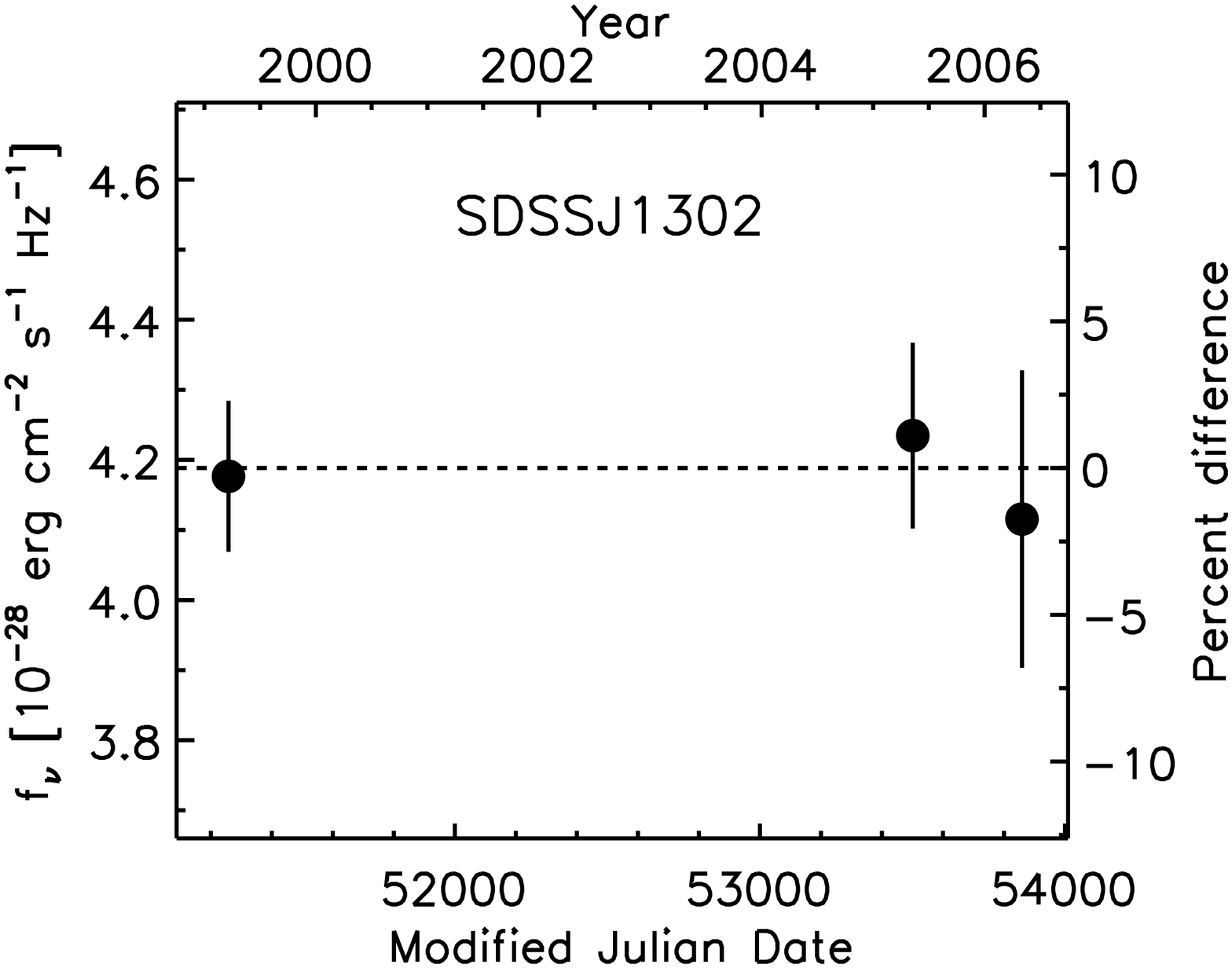}
\includegraphics[angle=90,scale=0.25]{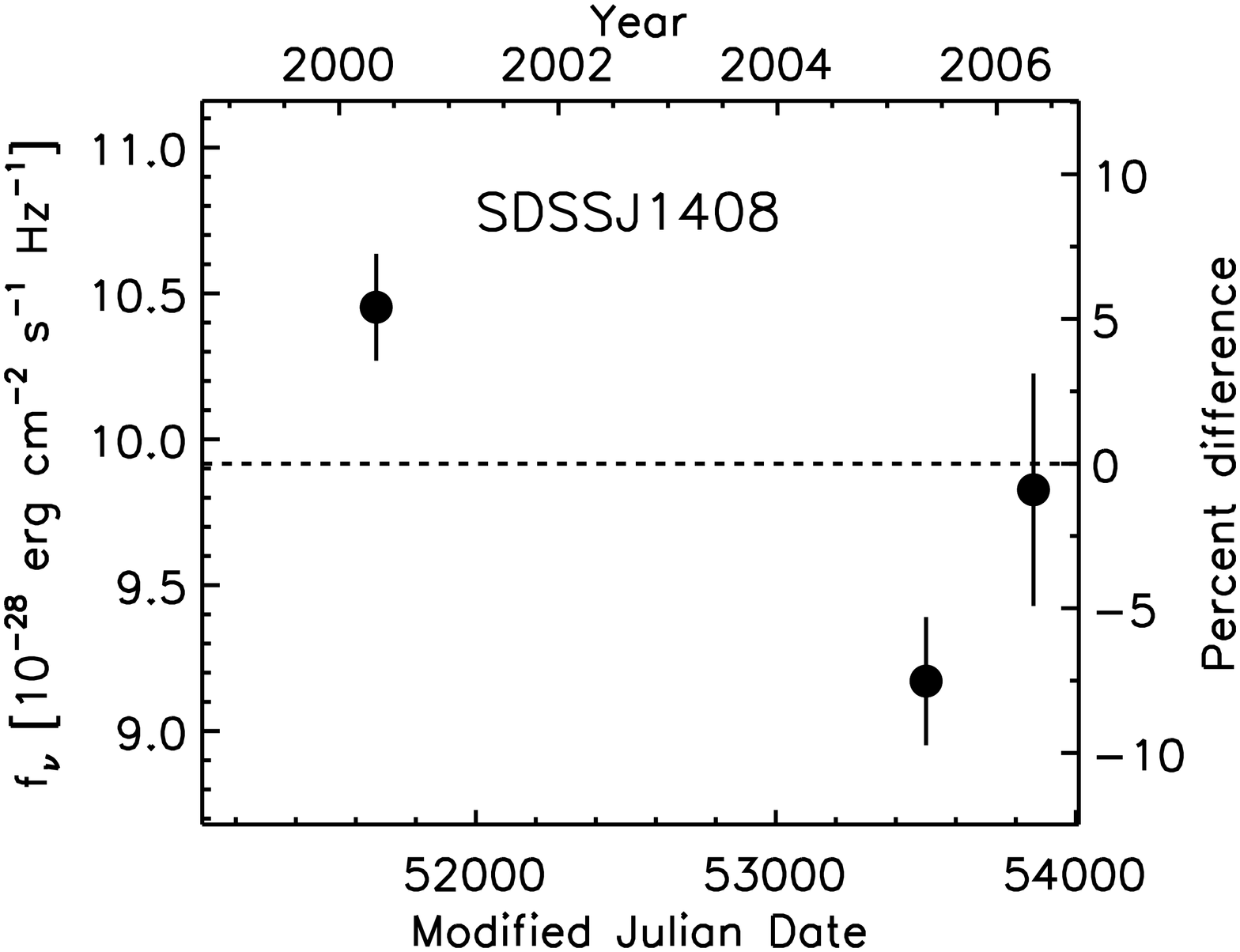}
\end{center}
\begin{center}
\includegraphics[angle=90,scale=0.25]{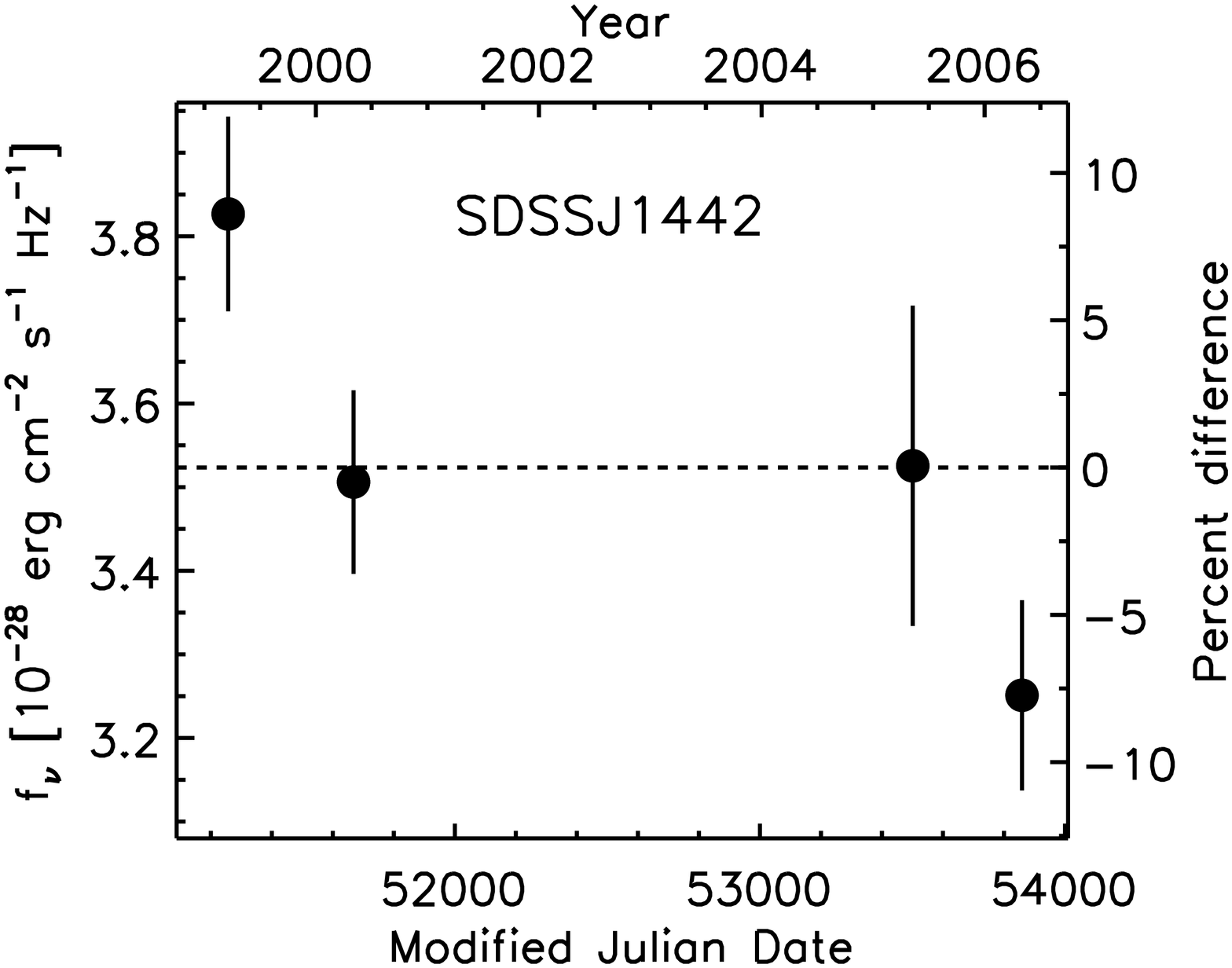}
\includegraphics[angle=90,scale=0.25]{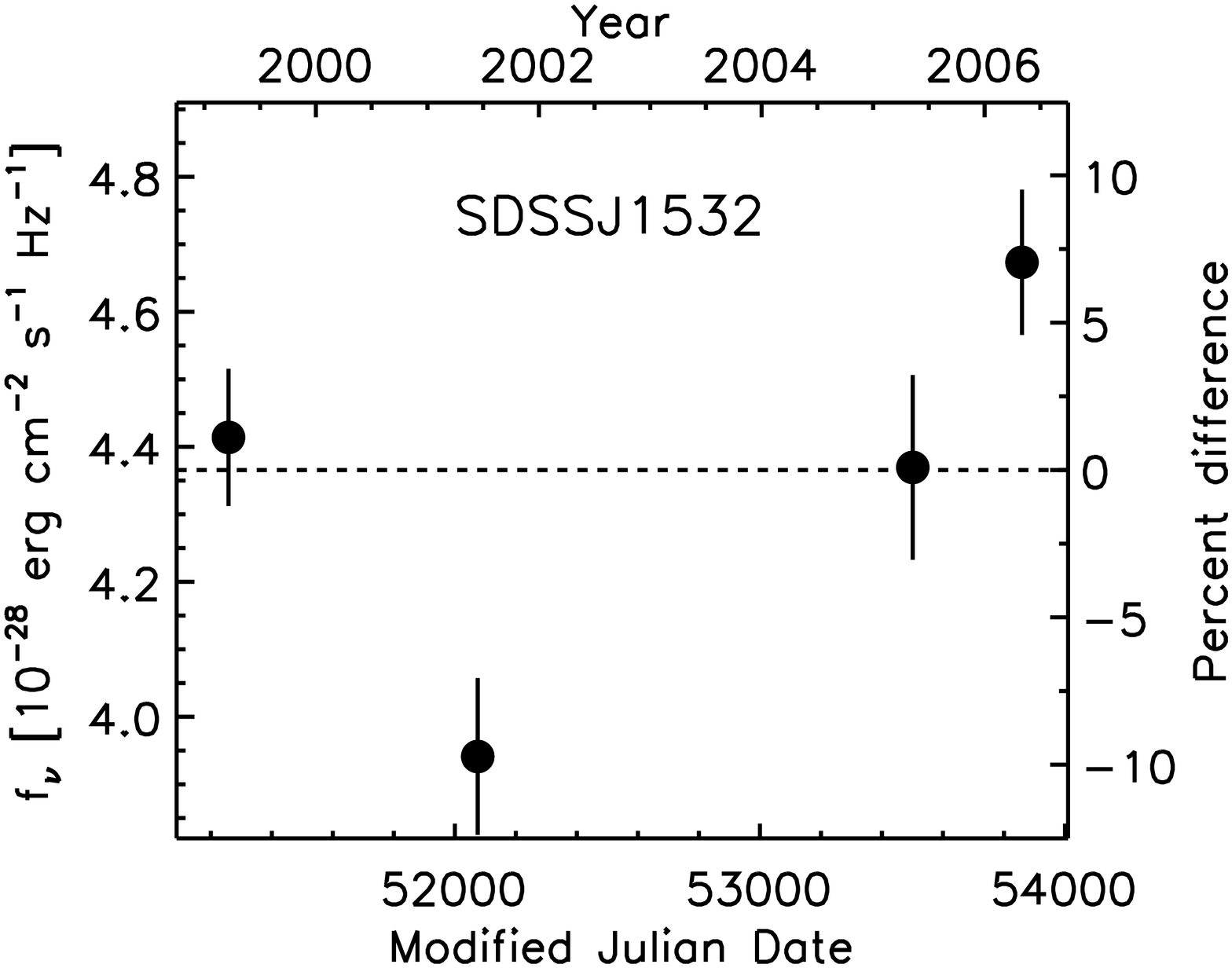}
\caption{Light curves from $i$-band photometry spanning a baseline of
at least six years for each WLQ.  The dotted line in each panel is the
weighted average of all the data points, which corresponds to
$i=19.84$ for SDSSJ1302, $i=18.91$ for SDSSJ1408, $i=20.03$ for
SDSSJ1442, and $i=19.80$ for SDSSJ1532.  The synthesized $i$-band
magnitudes from the spectroscopy epochs are also consistent with the
mean values, but they have larger uncertainties than the photometry
and we do not plot them for clarity.  The fluctuations around the mean
flux density are less than $\pm10$\% for all sources.}
\label{fig:lightcurves}
\end{center}
\end{figure*}

\section{Optical Variability}\label{sec:optical}

In this section, we present multiple epochs of optical spectroscopy
(Section~\ref{sec:spectra}) and $i$-band photometry
(Section~\ref{sec:phot}), designed to assess whether SDSSJ1302,
SDSSJ1408, SDSSJ1442, and SDSSJ1532 exhibit variability in line or
continuum emission that could explain the nature of their weak lines.

\subsection{Multi-Epoch Spectroscopy}\label{sec:spectra}

We obtained spectra for all four sources on 2004 July 7--8 with the
IMACS spectrograph \citep{big03} on the 6.5-m Baade Telescope at Las
Campanas Observatory.  Observations were made using a $1\arcsec$ slit
and a 600 line mm$^{-1}$ grism blazed at 7700~\AA.  The spectra were
dispersed onto two 2k$\times$4k chips, one that covered
5690--7660~\AA\ and another that covered 7720--10090~\AA.  The
spectral resolution ranged from $R\sim2500$ near $\lambda=6000$ \AA\
to $R\sim3500$ near $\lambda=9000$~\AA.  Total integration times
ranged from 60 minutes for SDSSJ1408 to 120 minutes for SDSSJ1532.

Another epoch of spectroscopy was acquired on 2005 May 8--9 with the
MARS spectrograph on the 4-m Mayall telescope at Kitt Peak.
Observations were carried out using a $1\arcsec$ slit with the
8050-450 grism (450 lines mm$^{-1}$, blazed at 8050 \AA), which
afforded wavelength coverage from 5400\AA\ to 1.1$\mu$m.  An OG-550
filter is built into this grating and blocks light shortward of
$\lambda=5500$\AA.  The resolution of the spectra is $R\sim750$.  The
total integration time was 40 minutes for SDSSJ1408 and 60 minutes
each for SDSSJ1302, SDSSJ1442, and SDSSJ1532.

SDSSJ1302 was also observed on 2002 January 12 with the Echelle
Spectrograph and Imager \citep[ESI,][]{shei02} on the Keck II
telescope for 15 minutes with $R\sim4000$.

We construct flat-field images using dome flats and generate
wavelength solutions for each object spectrum using lines identified
from HeNeAr lamp observations.  We combine separate exposures with
pixel shifts where appropriate and remove cosmic rays.  We generate
sensitivity functions using standard star spectra and apply the flux
calibration to the science targets.  One-dimensional spectra were
extracted using 1--2$\arcsec$ aperture sizes for the IMACS spectra and
from 3--4$\arcsec$ aperture sizes for the MARS spectra, using the
best-fit linear trace to the two-dimensional spectrum.  We make
corrections for telluric absorption based on extracted standard star
spectra.  The calibrated spectra are shown in
Figures~\ref{fig:j1302}--\ref{fig:j1532}, which also include the SDSS
spectra for SDSSJ1302, SDSSJ1408, and SDSSJ1442 obtained in
2000--2001, and the spectra of SDSSJ1532 previously published by
\citet{fan99} from the Double Imaging Spectrograph on the Apache Point
3.5-m telescope and the Low-Resolution Imaging Spectrograph (LRIS) on
the Keck II telescope.  We record the spectral slope and the EW of
\lya, as well as \sifour and \cfour in cases where the lines are
detected, for all spectra in Table~\ref{table:spectra}.

\subsection{Multi-Epoch Photometry}\label{sec:phot}

SDSS $i$-band photometry \citep{fuk96} was obtained for all four
sources on 2005 May 11 with OPTIC \citep{ton04} on the 3.5-m WIYN
telescope and on 2006 April 21--22 with 90Prime \citep{wil04} on the
Steward Observatory 2.3-m Bok telescope.  OPTIC consists of two
2k$\times$2k CCDs with a plate scale of 0.14$\arcsec$~pix$^{-1}$,
while 90Prime has four 4k$\times$4k CCDs with a plate scale of
0.45$\arcsec$~pix$^{-1}$.  The seeing during the observations ranged
from $1.2\arcsec$ to $2.0\arcsec$.

For the OPTIC data, we construct a night sky flat using a combination
of the science images, rejecting the five highest and five lowest
values for each pixel, and averaging the remaining values.  For the
90Prime data, we construct a flat-field image using 10 twilight sky
flats, rejecting the two highest and one lowest values of each pixel
and averaging the remaining seven pixel values.  The flat-field images
are scaled to their median value before we apply them to the data.

We perform aperture photometry on each science target and on nearby
SDSS stars in the field that are (1) brighter than the science target
and (2) faint enough to be in the regime where the CCD response is
linear to better than $1\%$ ($<30,000$ ADU).  These ``standard'' stars
are all fainter than $i=17$, have SDSS photometry that is accurate to
$\sim1\%$ \citep{ive04}, and are within 5$\arcmin$ of the science
target.  We use 2.8--4.2$\arcsec$ source apertures with a
4.9--5.6$\arcsec$ sky annulus for the OPTIC data and 2.7--5.4$\arcsec$
source apertures with a 6.8--9.0$\arcsec$ sky annulus for the 90Prime
data.  We measure the flux of the science target, with appropriate
aperture corrections, relative to each standard star, and then
calculate the weighted average of these measurements, which we quote
in Table~\ref{table:photometry} and use to construct light curves in
Figure~\ref{fig:lightcurves}.  We also include SDSS photometry in our
analysis, which was obtained in 1999--2001 (see
Table~\ref{table:photometry}).

\subsection{Variability, Weak Broad Lines, and Redshifts}\label{sec:optical_analysis}

Over the seven-year baseline of these observations, we find only
modest photometric and spectral variations.  As can be seen in
Figure~\ref{fig:lightcurves}, the $i$-band continuum emission
fluctuations are less than $\pm10$\% for all sources.  Similarly,
there is not much variability in the \lya, \siiv, and \cfour emission
lines, which remain weak in every spectrum in
Figures~\ref{fig:j1302}--\ref{fig:j1532}.  The most notable variation
can be seen in Figure~\ref{fig:j1442}, where SDSSJ1442 has a clear
detection of \lyalpha in all epochs, as well as a detection of \sifour
in the MARS spectrum.  The rest-frame EW of \lyalpha increases with
time, ranging from 9~\AA\ in the SDSS spectrum to 16~\AA\ in the MARS
spectrum, placing it outside our WLQ definition
($\textnormal{EW}<10$~\AA).  There are also detections of weak \sifour
and \cfour emission lines in the Keck spectrum of SDSSJ1302 (see
Figure~\ref{fig:j1302}) and in the IMACS spectrum of SDSSJ1408 (see
Figure~\ref{fig:j1408}).  These lines are not detected in the other
available spectra, which have lower S/N.

The lack of strong emission lines makes the task of measuring accurate
redshifts more difficult for WLQs than for normal quasars.  The
primary spectral feature in these sources is the break at the onset of
the \lyalpha forest, and we estimate the redshift for each WLQ by
assuming that this feature is at $\lrest=1216$~\AA.  These estimates
are consistent with previously published redshifts for SDSSJ1442
($z=4.51$) and SDSSJ1532 ($z=4.62$), both of which also have an LLS at
the same redshift (see Figures~\ref{fig:j1442} and \ref{fig:j1532}),
but they are different from the redshifts in the Fifth Data Release
Quasar Catalog for SDSSJ1302 (we find $z=4.48$ compared with
$z=4.4684$) and SDSSJ1408 (we find $z=3.95$ compared with $z=4.0075$).
The LLSs in the latter two spectra indicate somewhat lower redshifts
($z=4.37$ and $z=3.91$, respectively), but this is not necessarily a
problem because one does not always expect an LLS to be exactly at the
redshift of the quasar.

However, the weak emission lines detected in SDSSJ1302 (second panel
of Figure~\ref{fig:j1302}) are consistent with a lower redshift if we
interpret them as \sifour and \civ.  The central wavelengths of these
lines correspond to $z=4.38$ and $z=4.39$ or blueshifted velocities
$v=5500$~km~s$^{-1}$ and $v=4900$~km~s$^{-1}$ relative to $z=4.48$.
We also tentatively measure a \cfour blueshift in SDSSJ1408, but the
IMACS chip gap makes the central wavelength of \cfour highly
uncertain.  The \cfour emission line is typically blueshifted with
respect to the systemic quasar redshift \citep[e.g.,][]{gas82}, but
the amplitude of the velocity is $<1000$~km~s$^{-1}$ in most cases.
For example, \citet{ric02b} measured the velocity shift of \cfour
relative to \mgtwo for $\sim800$ quasars and found
$824\pm511$~km~s$^{-1}$ (mean and dispersion).  \citet{shen08} found
similar results, including very few \cfour blueshifts
$>3000$~km~s$^{-1}$, in a sample of $\sim15,000$ quasars.
\citet{ric02b} also found that the sources with larger \cfour
blueshifts have smaller EWs, so the interpretation that SDSSJ1302 has
high-ionization emission lines blueshifted by $\sim5000$~km~s$^{-1}$
is extreme, but perhaps feasible given the weakness of its lines.
Alternatively, the systemic redshift could be lower than our estimate
of $z=4.48$, implying that the line and continuum emission have been
strongly absorbed at $\lrest=1216$~\AA, which is the case for some
PDLAs with $z_{abs}\simeq z_{em}$ (see Section~\ref{sec:sample}).

\section{Polarization and Radio Continuum}\label{sec:synchrotron}

In this section we present optical polarimetry
(Section~\ref{sec:polarization}) and VLA observations
(Section~\ref{sec:radio}) of SDSSJ1302, SDSSJ1408, SDSSJ1442, and
SDSSJ1532 designed to assess whether they could be powered by
synchrotron emission from a relativistic jet, as is the case for BL
Lacs.

\subsection{Multi-Epoch Optical Polarimetry}\label{sec:polarization}

We obtained optical polarization measurements on 2005 May 13--16 and
2006 May 4--5 using the CCD Imaging/Spectropolarimeter
\citep[SPOL,][]{sch92a} on the Steward Observatory 2.3-m Bok
Telescope.  The instrument was used in imaging polarimetry mode, which
provides a $51\arcsec\times51\arcsec$ field of view with
$\sim0.5\arcsec$ pixels.  In this configuration, the key optical
elements are a rotatable $\lambda/2$ waveplate and a Wollaston prism,
which passes both orthogonal polarizations and splits them into
separate beams that are focused onto a $1200\times800$ SITe CCD
detector.  We obtained data using sequences of exposures that produce
two images for both beams, which can be analyzed to measure the {\it
Q} and {\it U} Stokes parameters.  We calibrated the data by (1)
inserting a NICOL prism into the beam, (2) observing an unpolarized
standard star, and (3) measuring polarized standard stars from
\citet{sch92b}.

In 2005 May, we used an $I$ filter; in 2006 May, we used an $R68$
filter, which is designed to transmit light redward of $6800$~\AA\ and
has $>90\%$ transmittance beyond $7200$~\AA.  The seeing during the
observations ranged from 1.5$\arcsec$ to 2.5$\arcsec$.  Additional
imaging polarimetry data for SDSSJ1532 were obtained on 1999 May 18
\citep[these data were published by][]{fan99} and 2000 January 8.
These observations used an $R72$ filter designed to transmit light
redward of $7200$~\AA.

We trim and bias-subtract the images using standard routines in IRAF.
We create a master flat-field image for both the {\it Q} and {\it U}
sequences by median combining the output of three {\it Q} dome-flat
sequences and three {\it U} dome-flat sequences.  After flat fielding,
we perform arithmetic operations on images from both beams to produce
{\it Q} and {\it U} images, as well an {\it I} Stokes image
corresponding to the total flux from the source.  We perform aperture
photometry on these images to measure the normalized Stokes parameters
$q=Q/I$ and $u=U/I$.  We used $3\arcsec$ apertures and a
5--10$\arcsec$ sky annulus for all the images.  Absolute flux
calibrations are not necessary for our polarization measurements, so
we make no effort to correct for source flux extending outside the
photometric aperture.

We calculate the linear polarization
$P_{\scriptsize{\textnormal{obs}}}=\sqrt{q^2+u^2}$ and the
polarization angle $\theta=0.5~\arctan(u/q)$ for each science target.
The results are shown in Table~\ref{table:polarization} and discussed
in Section~\ref{sec:nosynchrotron}.  We quote both the observed
polarization $P_{\scriptsize{\textnormal{obs}}}$ and the polarization
with a first-order correction for statistical bias \citep{war74},
$P=(P_{\scriptsize{\textnormal{obs}}}^2 - \sigma_{P}^2)^{1/2}$; the
latter value is used in our analysis.

\subsection{VLA Radio Continuum Data}\label{sec:radio}

We obtained observations with the VLA in the A-array configuration
\citep{tho80} on 2006 May 10.  All four sources were observed in the
$L$ band ($\lambda\sim20$~cm) with the L1 frequency combination, which
consists of $2\times50$~MHz intermediate frequencies (IFs) centered at
$\nu=1365$~MHz and $\nu=1435$~MHz.  To measure radio spectral slopes,
the two sources with FIRST radio detections (SDSSJ1408, SDSSJ1442)
were also observed in C-band ($\lambda\sim6$~cm) with $2\times50$~MHz
IFs centered at $\nu=4835$~MHz and $\nu=4885$~MHz .  There were 22
antennas in operation during the observations, and the longest antenna
baseline was $\sim36$~km, which provided an angular resolution of
$1.4\arcsec$ FWHM in the $L$ band and $0.4\arcsec$ in the $C$ band.
The total on-source integration times were 150 minutes for SDSSJ1532,
100 minutes for SDSSJ1302, and 25 minutes at both $L$ and $C$ bands
for SDSSJ1408 and SDSSJ1442, respectively.  Primary and secondary
calibrator sources were observed periodically, and absolute
calibration was tied to observations of 3C~286.  The data for our
science targets were calibrated using standard routines in AIPS.
Several antenna baselines were affected by RFI noise in the $L$ band,
and we flagged the $UV$ data with anomalous amplitudes.  The data were
cleaned and maps were made with the IMAGR task within AIPS.  The
morphologies of the two detected sources (SDSSJ1408, SDSSJ1442) are
consistent with point sources.  The flux from each was measured using
the IMEAN task with circular apertures at least twice the radius of
the beam.  The resulting flux densities are given in
Table~\ref{table:radio} and discussed in
Section~\ref{sec:nosynchrotron}.

\subsection{Optical Polarization, Radio Loudness, and Spectral Slope}\label{sec:nosynchrotron}

We do not detect any significant optical polarization from SDSSJ1302
in either 2005 May or 2006 May.  The $1\sigma$ uncertainty in this
measurement is $0.68\%$, and we set a $3\sigma$ upper limit of
$P<2.0\%$ for this WLQ.  We do detect a low level of polarization at
both epochs for SDSSJ1408 and SDSSJ1442, with weighted average values
of $P=1.50\%\pm0.37\%$ and $P=2.43\%\pm0.63\%$, respectively.  From
measurements of foreground stars, we estimate the value of
interstellar polarization along the line of sight to SDSSJ1408 and
SDSSJ1442 to be no greater than $\sim0.1\%$.  We detect polarization
from SDSSJ1532 in all four epochs of observation, although never with
$>3\sigma$ significance in a single epoch.  This source is at a fairly
low Galactic latitude, and we measure interstellar polarization
$P\sim1\%$.  Taking the weighted average of the $Q$ and $U$ values
from all four epochs of observations for SDSSJ1532, we find
$P=1.72\%\pm0.44\%$.  If we group the two older observations (1999 May
and 2000 January) separately from the two newer observations (2005 May
and 2006 May), we find $2.91\%\pm0.91\%$ and $1.30\%\pm0.50\%$
respectively.  Three of the WLQs in our full $z>3$ sample
(SDSSJ114153.34+021924.3, SDSSJ121221.56+534127.9, and
SDSSJ123743.08+630144.8) were observed by \citet{smi07}, who found
$P=0.90\%\pm0.64\%$, $P=1.32\%\pm0.58\%$, and $P=1.21\%\pm0.78\%$
respectively.  All of these values are below the nominal level for the
definition of highly polarized objects \citep[$P>3$\%, e.g.,][]{imp90}
and within the range for optically selected quasars without a
synchrotron component \citep[e.g.,][]{ber90}.

We do not detect SDSSJ1302 or SDSSJ1532 in our $L$-band radio
observations.  The maps for these two sources have $1\sigma$ noise
levels of 35~$\mu$Jy beam$^{-1}$ and 28~$\mu$Jy beam$^{-1}$,
respectively.  We calculate limits on the radio-loudness parameters
for these two sources using flux densities at rest-frame 6~cm
determined from the $3\sigma$ upper limits at observed-frame 20~cm,
assuming $\alpha_{r}=-0.8$, and flux densities at rest-frame 2500~\AA\
extrapolated between the observed frame $1.2$ and $1.6~\mu$m ($J$ and
$H$ bands) flux densities.  We find $R<2.5$ for SDSSJ1302 and $R<1.8$
for SDSSJ1532.  A flatter assumed radio spectral slope would place
even tighter limits of $R$ (e.g., for $\alpha_{r}=0$ the limits drop
to $R<1.7$ and $R<1.2$, respectively).  We measure flux densities for
SDSSJ1408 at the $L$ band and $C$ band of $0.89\pm0.07$~mJy and
$0.46\pm0.06$~mJy, corresponding to a radio spectral slope
$\alpha_r=-0.52$.  The $L$-band measurement is 25\% or $1.9\sigma$
smaller than the FIRST-integrated flux density (1.18~mJy, rms noise
0.14~mJy beam$^{-1}$), so there may be moderate variability, but it is
not highly significant.  The flux densities for SDSSJ1442 at the $L$
band and $C$ band are $0.83\pm0.07$~mJy and $0.45\pm0.06$~mJy,
corresponding to $\alpha_r=-0.49$.  The $L$-band measurement is 56\%
or $6.3\sigma$ smaller than the FIRST integrated flux density
(1.87~mJy, rms noise 0.15~mJy beam$^{-1}$), which indicates
significant variability. The values of the radio spectral slope are
both near the division between steep-spectrum and flat-spectrum radio
sources at $\alpha_r=-0.5$.  We calculate the radio-loudness
parameters for these two sources using their 20~cm flux densities,
their measured radio spectral slope, and an extrapolation between the
1.2 and 1.6~$\mu$m flux densities.  We find $R=10.0$ for SDSSJ1408 and
$R=29.3$ for SDSSJ1442.

\section{The Nature of Weak Emission-Line Quasars}\label{sec:discussion}

As shown in Section~\ref{sec:sample}, the line luminosities of
high-redshift WLQs are $4\times$ fainter than those of normal $z>3$
quasars, and their continuum luminosities are 40\% brighter.  There
are two hypotheses that are consistent with this result: (1) WLQs are
intrinsically less luminous than normal quasars in terms of both line
and continuum emission, but a relativistic jet beamed toward us
amplifies their continua; or (2) WLQs have the same intrinsic
continuum properties as normal quasars, but some physical process,
either a lack of line-emitting gas or obscuration along the line of
sight, causes the observed \lyalpha and other UV emission lines to be
weak.  In this section, we discuss how our results fit within these
two hypotheses, and what physical processes may be at work.

\subsection{Arguments Against Continuum Boosting}

\subsubsection{UV--IR Properties}

We find that all four WLQs with {\it Spitzer} photometry show emission
from hot ($T\sim1000$~K) dust.  This places strong constraints on any
continuum boosting because relativistic jets have no effect on thermal
dust emission.  We can thus rule out continuum boosting for the
radio-undetected WLQs (SDSSJ1302, SDSSJ1532), whose UV--IR SEDs
closely match those of typical quasars in all respects, including
their ratio of power law to thermal dust emission.  The radio-detected
sources (SDSSJ1408, SDSSJ1442) have weaker dust emission by factors of
1.5--1.9 (in the MIPS band) relative to the mean quasar SEDs, but this
is well within the factor of 2--3 scatter of normal type 1 quasar
SEDs.  Thus their UV--optical emission may be boosted by a factor of
$\sim2$, but no more, which is not sufficient to explain the extreme
weakness of their lines

Further evidence against continuum boosting comes from the fact that
WLQs do not exhibit strong optical polarization.  The low levels of
polarization observed in SDSSJ1408, SDSSJ1442, and SDSSJ1532 are
probably intrinsic, but are too small to imply that this polarization
comes from synchrotron emission, as is the case for BL Lacs.
Radio-selected BL Lacs are found to be highly polarized ($P>3\%$)
$\sim90\%$ of the time (i.e., in a single epoch, one would expect that
9/10 radio-selected BL Lacs would show $P>3\%$) and X-ray-selected BL
Lacs are found to be polarized roughly half of the time
\citep[e.g.,][]{jan94}.  The fact that no WLQ is found to be highly
polarized in several epochs of observations indicates that these
objects are significantly less polarized than even X-ray-selected BL
Lacs and consistent with the polarizations of normal quasars
\citep{ber90}.

There is also no evidence of strong optical variability in
WLQs.\footnote{\citet{sta05} measured a $\sim20$\% drop in flux in $R$
and $I$ bands for SDSSJ1532 between 2000 and 2001 and claimed more
dramatic fading relative to the SDSS $i$-band photometry in 1999.
Their results are called into question by the second epoch of SDSS
photometry in 2001 (see Table~\ref{table:photometry}), which shows
only a 10\% drop relative to 1999.  Their claim of a $\sim50$\% flux
difference between the 1999 SDSS $i$-band epoch and their 2000
$R$-band photometry is erroneous because they fail to account for the
fact that the $R$-band flux is heavily absorbed by the \lyalpha
forest.}  The fluctuations that are seen in
Figure~\ref{fig:lightcurves} are consistent with those of normal
quasars \citep[e.g.,][]{van04}.  We do see variability in both
\lyalpha EW and radio flux in SDSSJ1442, but the lack of any
corresponding optical continuum fluctuations in this source argues
against the continuum boosting scenario.  There is no evidence to
suggest that the increase in its line strength with time is a
reverberation effect, although our temporal sampling is not ideal
since the light crossing time of the broad-line region (BLR) is
expected to be $\sim5$~yr in the observed frame at this luminosity
\citep[e.g.,][]{lao98, ben09}.  The drop in the radio flux is
significant, and it is not required that the radio and optical
emission of a jet vary in a synchronized manner, but BL Lacs are found
to be variable at all wavelengths on a variety of timescales, so the
lack of any significant optical continuum variability indicates that
this emission is not likely coming from a jet.

\subsubsection{Radio Properties}

If the physical process causing weak line emission were continuum
boosting by relativistic jets, we would expect a large fraction of
WLQs to be radio loud.  Instead, we find no statistical difference
between the distribution of radio-loudness parameters for WLQs and
normal quasars.  The precise number of sources with $R>10$, a value
which is often used to describe radio loudness\footnote{A more
stringent definition is $R>100$, in which case objects with $10<R<100$
are referred to as radio-moderate.}, is not well constrained for
either normal quasars or WLQs because most have radio upper limits
that are still consistent with $R>10$.  The SDSS and FIRST data do
definitively indicate, however, that 92\% of WLQs and somewhere
between 93.3\% and 96.1\% of normal quasars have $R<40$, in contrast
to BL Lacs, the majority of which have $R>100$ \citep[see, e.g.,
Figure 5 of ][]{shem09}.  Almost all WLQs are either radio-quiet or
radio-moderate, indicating that if their continua are boosted, the
effect is monochromatic (i.e., roughly equal in the radio and
rest-frame UV) and distinct from the jet mechanism at work in BL Lacs.
\citet{shem09} discussed the potential association of WLQs with
radio-weak BL Lacs \citep[e.g.,][]{lon04, col05, and07, plo08} and
pointed out that the lack of typical (i.e., radio-loud) BL Lacs at
high redshift makes it difficult to connect the two phenomena.
Further evidence that we are not seeing pole-on radio jets in WLQs
comes from the $\alpha_{r}\sim-0.5$ radio spectral slopes for
SDSSJ1408 and SDSSJ1442, which are significantly steeper than the
typical slopes for BL Lacs, $\alpha_{r}\sim0.3$
\citep[e.g.,][]{sti91}.

\subsection{The Remaining Possibilities}

A variation on the continuum boosting hypothesis involves
gravitational lensing, where WLQs could either be (1) strongly lensed
galaxies or (2) normal quasars whose continuum emission has been
microlensed by a star in an intervening galaxy.  \citet{shem06} ruled
out the strongly lensed galaxy hypothesis for WLQs with strong X-ray
detections on the basis of their X-ray-to-optical flux ratios, which
are typical for quasars.  We additionally rule it out for all four
WLQs in Figure~\ref{fig:seds_template} because their UV--IR SEDs match
those of normal quasars.  The microlensing hypothesis has received
some attention in the literature, and several authors have used the
variability and EW distributions of quasars to put constraints on the
properties of lensing objects \citep[e.g.,][]{dal94,zac03,wie03}.  The
characteristic timescale for microlensing is $\sim10$~yr for a stellar
lens in a foreground galaxy \citep{gou95}, so we cannot rule out
microlensing for WLQs, but there is no evidence of fading continua
over 6--7~yr of observations.

The variety of arguments against continuum boosting as a cause for the
weak emission-line strength of WLQs, coupled with evidence against
lensing, implies that WLQs are a rare, unique population at high
redshift.  However, there are several objects at lower redshift whose
physical properties may be related.  One of these is the $z=0.94$
radio-quiet quasar PG 1407+265, whose properties are described in
detail by \citet{mcd95}.  This object has very weak \lyalpha emission
(\ew$=8$~\AA) and does not exhibit polarization \citep{ber90} or
optical--UV variability.  \citet{blu03} and \citet{gal06} observed a
factor of $\sim2$ variability in the radio and X-ray bands, similar to
the radio variability we see in SDSSJ1442
(Section~\ref{sec:nosynchrotron}), but this seems to be unrelated to
its optical--UV continuum flux.  It also has a weak \mgtwo emission
line ($\ew=24$~\AA), a somewhat stronger H$\alpha$ line
($\ew=126$~\AA), and unusually strong \fetwo emission.  Interestingly,
its weakly detected \cfour line ($\ew=4$~\AA) is blueshifted by
$\sim4000$~km~s$^{-1}$ with respect to Ly$\alpha$, similar to the
blueshift for SDSSJ1302 discussed in
Section~\ref{sec:optical_analysis}.  Other objects we are aware of
with weak \lyalpha emission and blueshifted \cfour lines are
SDSSJ152156.48+520238.4 \citep[$z=2.2$,][]{jus07} and HE 0141-3932
\citep[$z=1.8$,][]{rei05}.  \citet{ric02b} argued that observed \cfour
blueshifts are due to a lack of flux in the red wing of the emission
line, and their discussion of this phenomenon in the context of
cloud-based and accretion-disk-wind models for the BLR is pertinent
here.

Under the hypothesis that WLQs have normal quasar continuum
properties, but that some physical process causes the emission lines
to be weak, several of the possible interpretations for PG1407+265
mentioned by \citet{mcd95} are relevant for WLQs: (1) the BLR could
have anomalous properties or a low covering factor, (2) an exceptional
geometry could cause the BLR to see a continuum that is different from
the one that we see, or (3) the BLR could be covered by a patchy BAL
region that does not affect the continuum.  In the context of
explaining physical processes that would result in weak or absent
broad lines, \citet{nic03} discussed a scenario where the BLR forms
via accretion disk instabilities at a critical radius where gas
pressure begins to dominate over radiation pressure; at low accretion
rates / luminosities the BLR would move toward smaller radii and
eventually cease to exist.  Similarly, \citet{lao03} pointed out that
if the BLR cannot survive at line widths $\Delta
v>25,000$~km~s$^{-1}$, a minimum luminosity / accretion rate is
implied, and he discussed how the outer and inner boundaries of the
BLR may be set by suppression of line emission by dust and thermal
processes.  \citet{cze04} went a step further and calculated the
minimum radius, minimum Eddington ratio, and maximum line width for
which the BLR is expected to exist in ADAF and disk-evaporation
models.  However, the association of high-redshift WLQs with low
accretion rates, which would explain their weak lines in the above
theoretical scenarios, is ruled out by the fact that their continuum
properties (X-ray, UV, optical, IR, radio) are comparable to those of
normal quasars.

Another relevant object at low redshift ($z=0.192$) is the radio-quiet
quasar PHL~1811, a narrow-line Seyfert 1 galaxy (NLS1) with weak
\lyalpha and \cfour emission \citep{lei07b}, weak X-ray emission, and
a steep X-ray spectrum \citep[$\alpha_{ox}=-2.3$,
$\Gamma=2.3$,][]{lei07a}.  \citet{lei07b} showed that the weak
high-ionization emission lines in this quasar can be explained by its
soft SED in the sense that a lack of high-energy photons prevents
typical gas photoionization processes from occurring.  They speculated
that such weak X-ray and UV line emission may be associated with a
high accretion rate, which is often invoked as a physical
interpretation for NLS1s \citep[e.g.,][]{bol96}.  Similarly, the local
NLS1 NGC4051 ($z=0.002$) exhibited weak X-ray emission and a weak
$\heii~\lambda4686$ broad emission line (ionization potential 54.4~eV)
during the final months of a three-year monitoring campaign by
\citet{pet00}, while its broad H$\beta$ line remained strong.  The
authors speculated that the inner part of the accretion disk may have
become advection-dominated during this period, suppressing the X-ray
and far-UV continuum and also the high-ionization lines.

We do not have constraints on the strength of the low-ionization lines
(e.g., \mgii, Balmer lines) in WLQs, but further observations are
warranted to test where these lines are stronger than the
high-ionization lines, as is the case for PG 1407+265, PHL 1811, and
NGC 4051 (in its low X-ray flux state).  In models that describe the
BLR in terms of a disk wind \citep[e.g.,][]{mur95}, the low-ionization
lines are produced in the accretion disk and the high-ionization lines
are produced in the outflowing wind, so the UV emission lines could be
suppressed either by an abnormal photoionizing continuum or by a
process that prevents the disk wind itself from forming; either of
these could be associated with a high accretion rate.  Evidence of the
accretion rate being the physical driver of the Baldwin effect is
presented by \citet{bas04} and \citet{bac04}, but it is not clear if
such an inverse relationship between the \cfour EW and accretion rate
persists at higher luminosities.  \citet{shem09} explored the
possibility that WLQs could be extreme quasars with high accretion
rates, and correspondingly steep X-ray spectra, by jointly fitting the
available X-ray data; they found a slope that is consistent with those
of normal radio-quiet quasars, but higher-quality X-ray spectra are
required to test this hypothesis properly.

Finally, it is worth considering the effects of absorption on the
observed \lya+\nfive EWs of WLQs.  As discussed in
Section~\ref{sec:sample}, there is evidence of strong intervening
absorption in a fraction of the WLQ sample.  While we flag those WLQs
with obvious PDLA systems, absorption by lower column densities of
material could also affect the remainder of the sample.  In
Figure~\ref{fig:composite}, the decrease in the \lya/\nfive ratio and
the redward migration of the peak of the \lya+\nfive feature as one
moves toward lower EWs could be explained by absorption that affects
not just the blue side of \lya, but also the peak of the line and
emission redward of the peak.  Such behavior is also seen, although to
a lesser extent, in high-redshift quasar composite spectra presented
by \citet{die02} and \citet{fan04}.  There is no evidence of
corresponding \cfour absorption, however, so the nature of the
absorption would have to be different than in BALs and systems with
intrinsic narrow-line absorbers \citep[e.g.,][]{cre03}.  One scenario
that would explain the absorption of \lyalpha and not metal lines
would be the infall of pristine gas from the IGM
\citep[e.g.,][]{bar03}.  There is certainly evidence of strong IGM
\hone opacity blueward of \lya, and we cannot rule out absorption at
$\lrest\geq1216$~\AA\ in some WLQs, but it does not explain the \cfour
weakness, and we conclude that most WLQs are likely to have
intrinsically weak emission lines.

\section{Conclusions}\label{sec:conclusions}

We have identified a sample of 74 high-redshift ($z>3$) SDSS quasars
with weak UV emission lines and obtained IR, optical, and radio
observations for four such WLQs at $z>4$.  We find that the optical
continuum properties of WLQs are similar to those of normal quasars,
while their $\lya+\nfive$ line luminosities are significantly weaker,
by a factor of $\sim4$ on average.  The radio-loudness parameters of
WLQs are also statistically indistinguishable from those of normal
quasars.  The strong spectral break caused by \lyalpha forest
absorption makes it easier to optically select WLQs at higher
redshift, and the high WLQ fraction at $z>4.2$ ($\sim6$\% of quasars)
indicates that they may be quite common in general, although difficult
to select in most quasar surveys.  All WLQs with IR observations show
evidence of hot ($T\sim1000$ K) dust emission, which rules out
hypotheses that invoke relativistic continuum boosting to explain
their weak line emission.  Their $\lrest=0.1$--$5~\mu$m SEDs closely
resemble the mean quasar SED.  They are not highly variable or
polarized in the optical, and their weak emission lines persist
through several epochs of spectroscopy.  In addition, the radio
spectral slopes of radio-detected WLQs are also significantly steeper
than those of BL Lacs.

This evidence supports the interpretation that WLQs have the same
intrinsic continuum properties as normal quasars, but that some
physical process associated with the line-emitting gas results in weak
lines.  We are not able to discriminate among various possibilities
that could cause the exceptional emission-line weakness in WLQs, but
they include a low BLR covering factor, an abnormal geometry that
exploits the difference between our line of sight to the accretion
disk and the BLR, or suppression of high-ionization emission lines
perhaps related to a high accretion rate.  An important observational
test of the latter scenario will require near-IR spectroscopy to
search for the low-ionization emission line \mgii, as well as Balmer
lines, \feii, and \othree to facilitate comparison to the
\citet{bor92} Eigenvector 1 concept.  Further study is justified as
WLQs offer insight into emission-line physics and our overall
understanding of the quasar phenomenon.

\acknowledgments

We acknowledge helpful discussion with John Moustakas, Amelia Stutz,
and Brandon Kelly.  We thank George Rieke for comments on the
manuscript and Joseph Hennawi for providing the WIYN/OPTIC data.  We
thank the referee, Kirk Korista, for providing valuable feedback that
has resulted in an improved manuscript.  A.~M.~D and X.~F. acknowledge
support from NSF grants AST 03-07384 and AST 08-06861, a Packard
Fellowship for Science and Engineering, and a Guggenheim Fellowship.
W.~N.~B., O.~S., and D.~P.~S. acknowledge support from NASA LTSA grant
NAG5-13035.  M.~V. acknowledges support from grant HST-GO-10417 from
NASA through the Space Telescope Science Institute, which is operated
by the Association of Universities for Research in Astronomy, Inc.,
under NASA contract NAS5-26555.  The SDSS is managed by the
Astrophysical Research Consortium for the Participating
Institutions. The Participating Institutions are the American Museum
of Natural History, Astrophysical Institute Potsdam, University of
Basel, University of Cambridge, Case Western Reserve University,
University of Chicago, Drexel University, Fermilab, the Institute for
Advanced Study, the Japan Participation Group, Johns Hopkins
University, the Joint Institute for Nuclear Astrophysics, the Kavli
Institute for Particle Astrophysics and Cosmology, the Korean
Scientist Group, the Chinese Academy of Sciences (LAMOST), Los Alamos
National Laboratory, the Max-Planck-Institute for Astronomy (MPIA),
the Max-Planck-Institute for Astrophysics (MPA), New Mexico State
University, Ohio State University, University of Pittsburgh,
University of Portsmouth, Princeton University, the United States
Naval Observatory, and the University of Washington.

{\it Facilities:} \facility{Spitzer}, \facility{Bok},
\facility{Magellan:Baade}, \facility{ARC}, \facility{Mayall},
\facility{Keck:II}, \facility{WIYN}, \facility{VLA}.

\vskip1.5cm

\LongTables

\begin{deluxetable}{lrrrrrcrcc}
\tabletypesize{\scriptsize}
\tablecaption{Parameters for all $z>3$ Quasars\label{table:all}}
\tablewidth{0pt}
\tablehead{
\colhead{NAME (SDSS J)} & \colhead{\lya+\nv\tablenotemark{a}} & \colhead{$\sigma$} & \colhead{\civ\tablenotemark{a}} & \colhead{$\sigma$} & \colhead{$\alpha$\tablenotemark{b}} & \colhead{$z$}  & \colhead{$R$\tablenotemark{c}} &  \colhead{uniform\tablenotemark{d}} & \colhead{BAL\tablenotemark{e}}}
\startdata
$000046.41+011420.8$  &     87.2  &    3.2  &     71.4  &      2.6  &  -2.08  &  3.76  &   $< 18$  &  0  &  2 \\
$000051.56+001202.5$  &     23.8  &    1.3  &  \nodata  &  \nodata  &  -1.14  &  3.88  &   $  66$  &  0  &  2 \\
$000135.51-004206.7$  &     58.3  &    1.2  &  \nodata  &  \nodata  &   0.74  &  3.58  &   $< 17$  &  0  &  2 \\
$000221.11+002149.3$  &     23.9  &    0.5  &     16.5  &      1.5  &  -0.21  &  3.06  &   $ 132$  &  0  &  0 \\
$000238.41-101149.8$  &    117.3  &    2.9  &  \nodata  &  \nodata  &  -0.96  &  3.94  &   $< 28$  &  0  &  0 \\
$000252.72-000331.0$  &     62.7  &    4.7  &  \nodata  &  \nodata  &  -1.02  &  3.68  &   $  67$  &  0  &  0 \\
$000300.35+160027.6$  &     63.7  &    2.1  &     15.2  &      2.4  &  -1.18  &  3.68  &  \nodata  &  1  &  0 \\
$000303.35-105150.6$  &     89.0  &    1.2  &     79.8  &      2.1  &   0.03  &  3.65  &   $< 11$  &  0  &  3 \\
$000316.38-000732.4$  &     60.8  &    1.9  &     30.9  &      1.7  &  -0.08  &  3.18  &   $< 29$  &  0  &  0 \\
$000335.20+144743.6$  &     27.6  &    2.5  &  \nodata  &  \nodata  &  -0.17  &  3.48  &  \nodata  &  1  &  1 \\
\enddata
\tablecomments{Table~\ref{table:all} is published in its entirety in the
electronic edition of the journal.  A portion is 
shown here for guidance regarding its form and content.}
\tablenotetext{a}{\ew~[\AA].}
\tablenotetext{b}{$f_{\nu}\propto\nu^{\alpha}$.}
\tablenotetext{c}{Calculated as described in
Section~\ref{sec:radio_loudness}.  The upper limits use the FIRST
catalog detection limit, a $5\sigma$ threshold that includes CLEAN
bias.}

\tablenotetext{d}{Uniform selection flag in Fifth Data Release Quasar
Catalog: (0) not selected as a primary quasar target by the final
target selection algorithm given by \citet{ric02a}, (1) selected as a
primary quasar target by the final target selection algorithm.}

\tablenotetext{e}{BAL flag. (0) not identified as a BAL in either
\citet{tru06} or \citet{gib09} catalog, (1) identified as a BAL in
\citet{tru06} catalog only, (2) identified as a BAL in \citet{gib09}
catalog only, (3) identified as a BAL in both \citet{tru06} and
\citet{gib09} catalogs.}
\end{deluxetable}

\clearpage

\begin{deluxetable}{lrcrc}
\tabletypesize{\scriptsize}
\tablecaption{Sample of Weak Emission-Line Quasars\label{table:sample}}
\tablewidth{0pt}
\tablehead{
\colhead{NAME (SDSS J)} & \colhead{EW~(\AA)\tablenotemark{a}} & \colhead{$z$}  & \colhead{$R$\tablenotemark{b}} &  \colhead{uniform\tablenotemark{c}}}
\startdata
$004054.65-091526.7$                     &   4.5  &  4.98  &    $<16$  &  0 \\
$005421.42-010921.6$                     &  12.3  &  5.09  &    $<24$  &  0 \\
$010802.90-010946.1$                     &  12.3  &  3.37  &    $<14$  &  0 \\
$022337.76+003230.6$                     &   8.6  &  3.07  &   $  57$  &  0 \\
$025646.56+003858.3$                     &  12.8  &  3.47  &    $<12$  &  0 \\
$080523.32+214921.1$\tablenotemark{d,e}  &   5.6  &  3.46  &    $<10$  &  1 \\
$082059.35+561022.0$\tablenotemark{d,e}  &  12.2  &  3.64  &    $<23$  &  1 \\
$083122.57+404623.3$                     &   8.8  &  4.88  &    $<11$  &  1 \\
$084434.15+224305.2$\tablenotemark{d,e}  &  11.7  &  3.12  &    $<13$  &  1 \\
$091738.90+082053.9$\tablenotemark{d}    &  11.2  &  3.25  &    $<13$  &  1 \\
$093306.88+332556.6$                     &  11.9  &  4.56  &    $<16$  &  1 \\
$095108.76+314705.8$                     &   5.8  &  3.03  &   $ 127$  &  0 \\
$102949.80+605731.8$\tablenotemark{d}    &  14.0  &  3.19  &    $<22$  &  1 \\
$103240.53+501210.9$\tablenotemark{d}    &  13.9  &  3.82  &    $<12$  &  1 \\
$103601.12+084948.4$                     &   9.4  &  3.68  &    $<15$  &  1 \\
$104650.29+295206.8$                     &   6.3  &  4.27  &    $<14$  &  1 \\
$105049.28+441144.8$\tablenotemark{e}    &   5.3  &  4.32  &    $<13$  &  1 \\
$113354.89+022420.9$\tablenotemark{d}    &  15.0  &  3.99  &    $< 7$  &  0 \\
$113415.21+392826.1$                     &  13.8  &  4.83  &    $<29$  &  1 \\
$113729.42+375224.2$\tablenotemark{e}    &  14.7  &  4.17  &   $  59$  &  0 \\
$114153.34+021924.3$                     &   3.3  &  3.48  &   $  17$  &  0 \\
$114412.77+315800.8$                     &   7.0  &  3.23  &    $<11$  &  1 \\
$114434.60+510317.8$                     &   9.8  &  3.91  &    $<24$  &  1 \\
$114958.54+375115.0$                     &  13.3  &  4.31  &    $<34$  &  1 \\
$115254.97+150707.7$\tablenotemark{d}    &  12.8  &  3.33  &    $< 6$  &  1 \\
$115308.45+374232.1$\tablenotemark{d}    &  15.2  &  3.03  &    $<29$  &  1 \\
$115906.52+133737.7$\tablenotemark{d}    &  11.9  &  3.98  &    $< 2$  &  0 \\
$115933.53+054141.6$\tablenotemark{d}    &  10.5  &  3.29  &    $<20$  &  1 \\
$120059.68+400913.1$\tablenotemark{e}    &   6.9  &  3.37  &   $  64$  &  1 \\
$121221.56+534127.9$                     &   8.2  &  3.10  &    $< 5$  &  1 \\
$121812.39+444544.5$                     &  12.6  &  4.52  &    $<19$  &  1 \\
$122021.39+092135.8$\tablenotemark{d,e}  &  12.9  &  4.11  &    $<11$  &  1 \\
$122359.35+112800.1$\tablenotemark{d,e}  &  11.5  &  4.12  &    $<18$  &  1 \\
$122445.27+375921.3$                     &  15.1  &  4.30  &    $<18$  &  1 \\
$123116.08+411337.3$                     &  10.7  &  3.84  &    $<18$  &  1 \\
$123132.37+013814.0$                     &   6.3  &  3.23  &   $  84$  &  0 \\
$123315.94+313218.4$                     &  12.9  &  3.22  &   $ 106$  &  1 \\
$123540.19+123620.7$\tablenotemark{d}    &  10.5  &  3.21  &    $<25$  &  1 \\
$123743.08+630144.8$                     &   4.6  &  3.42  &    $< 7$  &  0 \\
$124204.27+625712.1$                     &  12.2  &  3.32  &   $  28$  &  1 \\
$125306.73+130604.9$\tablenotemark{d}    &  13.9  &  3.63  &    $< 7$  &  1 \\
$125319.10+454152.8$                     &  11.3  &  3.53  &    $<19$  &  1 \\
$130216.13+003032.1$                     &   7.8  &  4.47  &    $<23$  &  0 \\
$130332.42+621900.3$                     &  10.9  &  4.66  &    $<23$  &  1 \\
$131429.00+494149.0$\tablenotemark{d}    &  13.4  &  3.81  &    $< 8$  &  1 \\
$132603.00+295758.1$                     &  10.1  &  3.77  &    $< 7$  &  1 \\
$133146.20+483826.3$\tablenotemark{e}    &  15.3  &  3.74  &    $<11$  &  1 \\
$133422.63+475033.5$                     &  11.6  &  4.95  &    $<16$  &  1 \\
$134453.51+294519.6$                     &   9.1  &  4.71  &    $<14$  &  1 \\
$134521.39+281822.2$                     &   6.3  &  4.08  &    $<24$  &  1 \\
$140300.22+432805.3$                     &  14.5  &  4.70  &    $<21$  &  1 \\
$140850.91+020522.7$                     &   7.0  &  4.01  &   $  14$  &  0 \\
$141209.96+062406.8$                     &   2.4  &  4.47  &   $ 771$  &  1 \\
$141318.86+450522.9$\tablenotemark{d,e}  &   3.4  &  3.11  &   $1782$  &  1 \\
$142103.83+343332.0$\tablenotemark{e}    &   7.0  &  4.91  &    $<11$  &  1 \\
$142144.98+351315.4$\tablenotemark{d,e}  &   3.1  &  4.56  &    $< 9$  &  1 \\
$142257.66+375807.3$\tablenotemark{d,e}  &   8.2  &  3.16  &    $<21$  &  1 \\
$143009.55+550535.0$                     &   7.0  &  3.72  &    $<17$  &  1 \\
$144127.65+475048.8$\tablenotemark{d,e}  &  12.3  &  3.19  &    $< 7$  &  1 \\
$144231.71+011055.3$                     &  11.4  &  4.51  &   $  25$  &  0 \\
$150220.46+465233.5$                     &  11.4  &  4.26  &    $<18$  &  1 \\
$150739.67-010911.0$                     &  12.9  &  3.14  &    $<15$  &  1 \\
$152200.14+413741.7$                     &   9.3  &  3.24  &   $  17$  &  0 \\
$155203.30+352440.4$                     &   4.6  &  3.04  &    $< 8$  &  1 \\
$155645.31+380752.7$                     &  14.9  &  3.32  &    $<13$  &  1 \\
$160336.64+350824.2$                     &   5.3  &  4.46  &    $< 6$  &  1 \\
$161122.44+414409.6$\tablenotemark{d,e}  &   8.6  &  3.13  &    $< 7$  &  1 \\
$163411.82+215325.0$                     &   6.7  &  4.53  &    $<10$  &  1 \\
$210216.52+104906.6$\tablenotemark{d}    &  14.7  &  4.18  &  \nodata  &  1 \\
$214753.29-073031.3$                     &  14.5  &  3.15  &   $  18$  &  0 \\
$223827.17+135432.6$\tablenotemark{d,e}  &   1.2  &  3.52  &  \nodata  &  1 \\
$225246.43+142525.8$                     &  11.1  &  4.90  &  \nodata  &  1 \\
$233255.71+141916.4$                     &  10.9  &  4.75  &  \nodata  &  1 \\
$233446.40-090812.2$\tablenotemark{d}    &  12.6  &  3.32  &   $ 102$  &  0 \\
\enddata
\tablenotetext{a}{\ew(\lya+\nv)}
\tablenotetext{b}{Calculated as described in Section~\ref{sec:radio_loudness}. The upper limits use the FIRST catalog detection limit, a $5\sigma$ threshold that includes CLEAN bias.}
\tablenotetext{c}{Uniform selection flag in Fifth Data Release Quasar Catalog: (0) not selected as a primary quasar target by the final target selection algorithm given by \citet{ric02a}, (1) selected as a primary quasar target by the final target selection algorithm.}
\tablenotetext{d}{\ew(\civ$)>10$~\AA.}
\tablenotetext{e}{In \citet{pro08} catalog of Proximate Damped \lyalpha Absorbers.}
\end{deluxetable}

\vskip0.5cm

\begin{deluxetable}{ccccccccccc}
\tabletypesize{\scriptsize} 
\tablecaption{{\it Spitzer} Photometry\label{table:spitzer_photometry}} 
\tablewidth{0pt}
\tablehead{ \colhead{Name} & \colhead{3.6 [$\mu$Jy]} &
\colhead{$\sigma$} & \colhead{4.5 [$\mu$Jy]} & \colhead{$\sigma$} &
\colhead{5.8 [$\mu$Jy]} & \colhead{$\sigma$} & \colhead{8.0 [$\mu$Jy]}
& \colhead{$\sigma$} & \colhead{24.0 [$\mu$Jy]} & \colhead{$\sigma$} }
\startdata 
SDSSJ1302 & 78.3 & 3.9 & 62.6 & 3.1 & 62.9 & 3.1 & 84.6  & 4.2 & 470 & 24 \\ 
SDSSJ1408 & 86.4 & 4.3 & 76.3 & 3.8 & 80.5 & 4.0 & 140.6 & 7.0 & 575 & 29 \\ 
SDSSJ1442 & 37.9 & 1.9 & 33.0 & 1.7 & 35.2 & 2.7 & 67.8  & 3.7 & 191 & 24 \\ 
SDSSJ1532 & 83.5 & 4.2 & 70.8 & 3.5 & 72.1 & 3.6 & 99.8  & 5.0 & 534 & 27 \\ 
\enddata
\end{deluxetable}

\vskip0.5cm

\begin{deluxetable}{lllrrrrrr}
\tabletypesize{\scriptsize}
\tablecaption{Near-IR Photometry\label{table:nir_photometry}}
\tablewidth{0pt}
\tablehead{
\colhead{Name} & \colhead{Telescope/Instrument} & \colhead{Date} & \colhead{$J$~[$\mu$Jy]} & \colhead{$\sigma$} & \colhead{$H$~[$\mu$Jy]} & \colhead{$\sigma$} & \colhead{$K$~[$\mu$Jy]} & \colhead{$\sigma$} }
\startdata
SDSSJ1302 & Bok 2.3m/$256\times$256 & 2005 Apr 26 &  59.4 & 12.1 &  67.7 & 13.1 &\nodata   &\nodata  \\
SDSSJ1302 & Magellan/PANIC          & 2005 Apr 30 &\nodata   &\nodata  &\nodata   &\nodata  &  59.7 &  6.0 \\
SDSSJ1302 & APO 3.5m/NICFPS         & 2005 Jun 30  &  54.3 &  7.0 &  69.6 & 10.2 &  58.6 &  8.6 \\
SDSSJ1302 & Weighted average        &\nodata           &  56.3 &  9.3 &  68.9 & 11.4 &  59.1 &  7.8 \\
\\					        
SDSSJ1408 & Bok 2.3m/$256\times$256 & 2005 Apr 27 & 105.3 &  9.9 & 101.5 & 17.1 &\nodata   &\nodata  \\
SDSSJ1408 & Magellan/PANIC          & 2005 May 1    &\nodata   &\nodata  &\nodata   &\nodata  &  92.0 &  5.2 \\
SDSSJ1408 & APO 3.5m/NICFPS         & 2005 Jun 30  & 113.4 &  8.5 &  94.3 & 10.6 &  92.0 &  6.9 \\
SDSSJ1408 & Weighted average        &\nodata           & 111.3 &  9.4 &  95.1 & 11.6 &  92.0 &  6.8 \\
\\					        
SDSSJ1442 & Bok 2.3m/$256\times$256 & 2005 Apr 27 &  38.2 &  7.4 &\nodata   &\nodata  &\nodata   &\nodata  \\
SDSSJ1442 & Magellan/PANIC          & 2005 May 1    &\nodata   &\nodata  &\nodata   &\nodata  &  34.4 &  2.9 \\
SDSSJ1442 & APO 3.5m/NICFPS         & 2005 Jun 30  &  32.1 &  5.3 &  38.6 &  6.4 &  34.7 &  5.1 \\
SDSSJ1442 & Weighted average        &\nodata           &  34.2 &  6.3 &  38.6 &  6.4 &  34.7 &  4.1 \\
\\					        
SDSSJ1532 & Bok 2.3m/$256\times$256 & 2005 Apr 27 &  64.0 &  8.3 &\nodata   &\nodata  &\nodata   &\nodata  \\
SDSSJ1532 & APO 3.5m/NICFPS         & 2005 Jun 30  &\nodata   &\nodata  &  80.6 &  6.7 &  66.1 &  7.3 \\
SDSSJ1532 & Weighted average        &\nodata           &  64.0 &  8.3 &  80.6 &  6.7 &  66.1 &  7.3 \\
\enddata
\end{deluxetable}

\clearpage

\begin{deluxetable}{llllccc}
\tabletypesize{\scriptsize} 
\tablecaption{Spectroscopic Parameters\label{table:spectra}}
\tablewidth{0pt} 
\tablehead{ 
\colhead{Object} & \colhead{Instrument} &
\colhead{Date} & \colhead{$\alpha$} & \colhead{\lya} &
\colhead{\siiv} & \colhead{\civ} \\  
 & & & & \ewrest~[\AA] & \ewrest~[\AA] & \ewrest~[\AA]}
\startdata 
SDSSJ1302 & SDSS & 2001 Mar 26 & $-0.45$ & 2.5   &\nodata   &\nodata   \\ 
SDSSJ1302 & Keck/ESI  & 2002 Jan 12 & $-0.61$ & 2.4  & 2.5   & 4.7   \\
SDSSJ1302 & IMACS & 2004 Jul 8  & $-0.49$ & 2.3  &\nodata   &\nodata   \\ 
SDSSJ1302 & MARS  & 2005 May 9  & $-0.49$ & 2.3  &\nodata   &\nodata   \\ 
SDSSJ1408 & SDSS  & 2001 Mar 25 & $-0.57$ & 7.0  &\nodata   &\nodata   \\ 
SDSSJ1408 & IMACS & 2004 Jul 7  & $-0.55$ & 2.0  & 7.4   & 6.0   \\ 
SDSSJ1408 & MARS  & 2005 May 9  & $-0.55$ & 6.1  &\nodata   &\nodata   \\ 
SDSSJ1442 & SDSS  & 2000 Apr 29 & $-0.23$ & 9.1  &\nodata   &\nodata   \\ 
SDSSJ1442 & IMACS & 2004 Jul 7  & $-0.33$ & 12.9 &\nodata   &\nodata   \\ 
SDSSJ1442 & MARS  & 2005 May 9  & $-0.15$ & 15.7 & 7.4   &\nodata   \\ 
SDSSJ1532 & APO   & 1999 Mar 13 & $-1.50$ & 1.1  &\nodata   &\nodata   \\ 
SDSSJ1532 & KECK/LRIS  & 1999 May 13 & $-0.96$ & 1.2  &\nodata   &\nodata   \\ 
SDSSJ1532 & IMACS & 2004 Jul 8  & $-0.72$ & 2.5  &\nodata   &\nodata   \\ 
SDSSJ1532 & MARS  & 2005 May 8  & $-1.00$ & 1.1  &\nodata   &\nodata   \\ 
\enddata 
\end{deluxetable}

\vskip0.5cm

\begin{deluxetable}{lllrr}
\tabletypesize{\scriptsize} 
\tablecaption{SDSS $i$-band Photometry\label{table:photometry}}
\tablewidth{0pt} 
\tablehead{ \colhead{Name} & \colhead{Instrument} &
\colhead{Date} & \colhead{$f_{\nu}$ [$\mu$Jy]} & \colhead{$\sigma$ [$\mu$Jy]} } 
\startdata
SDSSJ1302 & SDSS        & 1999 Mar 21 &  41.8 & 1.1 \\ 
SDSSJ1302 & WIYN/OPTIC  & 2005 May 11 &  42.4 & 1.3 \\ 
SDSSJ1302 & Bok/90Prime & 2006 Apr 21 &  41.2 & 2.1 \\ 
\\
SDSSJ1408 & SDSS        & 2000 May 4  & 104.5 & 1.8 \\ 
SDSSJ1408 & WIYN/OPTIC  & 2005 May 11 &  91.7 & 2.2 \\ 
SDSSJ1408 & Bok/90Prime & 2006 Apr 21 &  98.3 & 4.0 \\ 
\\
SDSSJ1442 & SDSS        & 1999 Mar 20 &  38.3 & 1.2 \\ 
SDSSJ1442 & SDSS        & 2000 May 4  &  35.1 & 1.1 \\ 
SDSSJ1442 & WIYN/OPTIC  & 2005 May 11 &  35.3 & 1.9 \\ 
SDSSJ1442 & Bok/90Prime & 2006 Apr 21 &  32.5 & 1.1 \\ 
\\
SDSSJ1532 & SDSS        & 1999 Mar 21 &  44.1 & 1.0 \\ 
SDSSJ1532 & SDSS        & 2001 Jun 15 &  39.4 & 1.2 \\ 
SDSSJ1532 & WIYN/OPTIC  & 2005 May 11 &  43.7 & 1.4 \\ 
SDSSJ1532 & Bok/90Prime & 2006 Apr 21 &  46.7 & 1.1 \\ 
\enddata
\end{deluxetable}

\clearpage

\begin{deluxetable}{llrrrrrrrrr}
\tabletypesize{\scriptsize}
\tablecaption{Polarization \label{table:polarization}}
\tablewidth{0pt}
\tablehead{
\colhead{Name} & \colhead{Date} & \colhead{$q$} & \colhead{$\sigma_{\scriptsize{q}}$} & \colhead{$u$} & \colhead{$\sigma_{\scriptsize{u}}$} & \colhead{$P_{\scriptsize{\textnormal{obs}}}$~\tablenotemark{a}} & \colhead{$P$~\tablenotemark{b}} & \colhead{$\sigma_{\scriptsize{P}}$} & \colhead{$\theta$~\tablenotemark{c}} & \colhead{$\sigma_{\scriptsize{\theta}}$~\tablenotemark{d}} }
\startdata
SDSSJ1302                   & 2005 May 13 & 0.67  & 1.09 & -0.08 & 1.07 & 0.68 &\nodata       & 1.09 & 176.7 &\nodata  \\ 
SDSSJ1302                   & 2006 May 4  & 0.11  & 0.96 & -0.49 & 0.77 & 0.50 &\nodata       & 0.78 & 141.0 &\nodata  \\
SDSSJ1302                   &\nodata    & 0.36  & 0.72 & -0.35 & 0.63 & 0.50 & \bf{--}   & 0.68 & 157.8 &\nodata  \\
\\	                    
SDSSJ1408                   & 2005 May 13 & 2.05  & 0.84 & -0.19 & 0.72 & 2.06 & 1.88      & 0.84 & 177.4 & 12.7 \\
SDSSJ1408                   & 2006 May 4  & 0.87  & 0.41 & 0.83  & 0.38 & 1.21 & 1.14      & 0.40 & 21.9  & 10.0 \\
SDSSJ1408~\tablenotemark{e} &\nodata& 0.00  & 0.12 & -0.04 & 0.11 & 0.04 &\nodata       & 0.11 & 132.6 &\nodata  \\
SDSSJ1408~\tablenotemark{f} &\nodata    & 1.42  & 0.38 & 0.59  & 0.34 & 1.54 & \bf{1.50} & 0.37 & 89.7  & 5.7  \\  
\\ 	                    
SDSSJ1442                   & 2005 May 13 & 1.02  & 1.26 & 1.85  & 1.20 & 2.11 & 1.73      & 1.21 & 30.6  & 20.1 \\
SDSSJ1442                   & 2006 May 4  & 1.33  & 0.75 & 2.19  & 0.72 & 2.57 & 2.46      & 0.73 & 29.3  & 8.5  \\
SDSSJ1442~\tablenotemark{e} &\nodata    & -0.08 & 0.01 & 0.11  & 0.01 & 0.13 & 0.13      & 0.01 & 62.5  & 2.8  \\
SDSSJ1442~\tablenotemark{f} &\nodata    & 1.54  & 0.64 & 1.99  & 0.62 & 2.51 & \bf{2.43} & 0.63 & 26.1  & 7.4  \\
\\	                    
SDSSJ1532                   & 1999 May 18 & 1.45  & 1.13 & 2.23  & 1.05 & 2.66 & 2.44      & 1.08 & 28.5  & 12.7 \\
SDSSJ1532                   & 2000 Jan 8  & -1.33 & 1.65 & 3.48  & 1.75 & 3.73 & 3.30      & 1.74 & 55.4  & 15.1 \\
SDSSJ1532~\tablenotemark{e} &\nodata    & -0.83 & 0.01 & 0.00  & 0.01 & 0.83 & 0.83      & 0.01 & 89.9  & 0.4  \\
SDSSJ1532~\tablenotemark{f} &\nodata    & 1.37  & 0.93 & 2.72  & 0.90 & 3.05 & \bf{2.91} & 0.91 & 31.6  & 9.0  \\ 
\\
SDSSJ1532                   & 2005 May 13 & 0.46  & 1.35 & 3.88  & 1.55 & 3.90 & 3.59      & 1.54 & 41.6  & 12.4 \\
SDSSJ1532                   & 2006 May 4  & -0.09 & 0.49 & 1.08  & 0.53 & 1.08 & 0.94      & 0.53 & 47.4  & 16.1 \\
SDSSJ1532~\tablenotemark{e} &\nodata    & -0.99 & 0.13 & 0.39  & 0.14 & 1.06 & 1.06      & 0.13 & 79.3  & 3.7  \\
SDSSJ1532~\tablenotemark{f} &\nodata    & 0.95  & 0.48 & 1.01  & 0.52 & 1.39 & \bf{1.30} & 0.50 & 23.4  & 11.1 \\
\\
SDSSJ1532~\tablenotemark{g} &\nodata    & 1.04  & 0.43 & 1.44  & 0.45 & 1.78 & \bf{1.72} & 0.44 & 27.1  & 7.4  \\ 
\enddata
\tablecomments{The data in columns $3-9$ are all percentages.}
\tablenotetext{a}{~$P_{\scriptsize{\textnormal{obs}}}=\sqrt{q^2 + u^2}$}
\tablenotetext{b}{~$P=\sqrt{P_{\scriptsize{\textnormal{obs}}}^2 - \sigma_{P}^2}$}
\tablenotetext{c}{~$\theta=0.5~\arctan(u/q)$}
\tablenotetext{d}{~$\sigma_{\theta}=28.65~(\sigma_{P}/P)$}
\tablenotetext{e}{~Measurement of interstellar polarization.}
\tablenotetext{f}{~Weighted average corrected for interstellar polarization.}
\tablenotetext{g}{~Weighted average of four epochs corrected for interstellar polarization.}
\end{deluxetable}

\vskip0.5cm

\begin{deluxetable}{ccccccr}
\tabletypesize{\scriptsize} 
\tablecaption{{\it VLA} Continuum Observations\label{table:radio}}
\tablewidth{0pt}
\tablehead{
\colhead{Name} & \colhead{L band} & \colhead{rms noise} & \colhead{C band} & \colhead{rms noise} & \colhead{$\alpha$} & \colhead{$R$} \\
               & [mJy]            & [mJy/beam]          &  [mJy]           & [mJy/beam]          &                    &      }
\startdata
SDSSJ1302 &\nodata      & 0.035   &\nodata     &\nodata    &\nodata    & $<2.5$ \\
SDSSJ1408 & 0.891    & 0.074   & 0.461   & 0.060  & $-0.52$  & 10.1 \\
SDSSJ1442 & 0.830    & 0.066   & 0.446   & 0.060  & $-0.49$  & 29.3 \\
SDSSJ1532 & \nodata      & 0.028   &\nodata     &\nodata    &\nodata     & $<1.8$ \\
\enddata
\end{deluxetable}

\end{document}